\documentclass[final,3p,times,authoryear]{elsarticle}
\usepackage{hyperref}
\usepackage{graphicx}
\usepackage[T1]{fontenc}
\usepackage{endnotes}
\usepackage{setspace}

\usepackage{ifthen}

\usepackage{amssymb}

\journal{Physics Reports}

\def\ltsima{$\; \buildrel < \over \sim \;$}
\def\simlt{\lower.5ex\hbox{\ltsima}}
\def\gtsima{$\; \buildrel > \over \sim \;$}
\def\simgt{\lower.5ex\hbox{\gtsima}}

\def\gsim{~\rlap{$>$}{\lower 1.0ex\hbox{$\sim$}}}
\def\lsim{~\rlap{$<$}{\lower 1.0ex\hbox{$\sim$}}}

\newcounter{AGNDone}
\def\AGN{\ifthenelse{\equal{\arabic{AGNDone}}{0}}{active galactic nuclei (AGN)\setcounter{AGNDone}{1}}{AGN}}
\newcounter{CDMDone}
\def\CDM{\ifthenelse{\equal{\arabic{CDMDone}}{0}}{cold dark matter (CDM)\setcounter{CDMDone}{1}}{CDM}}
\newcounter{CMBDone}
\def\CMB{\ifthenelse{\equal{\arabic{CMBDone}}{0}}{cosmic microwave background (CMB)\setcounter{CMBDone}{1}}{CMB}}
\newcounter{ICMDone}
\def\ICM{\ifthenelse{\equal{\arabic{ICMDone}}{0}}{intracluster medium (ICM)\setcounter{ICMDone}{1}}{ICM}}
\newcounter{ISMDone}
\def\ISM{\ifthenelse{\equal{\arabic{ISMDone}}{0}}{interstellar medium (ISM)\setcounter{ISMDone}{1}}{ISM}}
\newcounter{NFWDone}
\def\NFW{\ifthenelse{\equal{\arabic{NFWDone}}{0}}{Navarro-Frenk-White (NFW)\setcounter{NFWDone}{1}}{NFW}}
\newcounter{SMBHDone}
\def\SMBH{\ifthenelse{\equal{\arabic{SMBHDone}}{0}}{supermassive black hole (SMBH)\setcounter{SMBHDone}{1}}{SMBH}}
\def\SMBHs{\ifthenelse{\equal{\arabic{SMBHDone}}{0}}{supermassive black holes (SMBH)\setcounter{SMBHDone}{1}}{SMBHs}}

\def\aj{AJ}                   
\def\araa{ARA\&A}             
\def\apj{ApJ}                 
\def\apjl{ApJ}                
\def\apjs{ApJS}               
\def\aap{A\&A}                
\def\aapr{A\&A~Rev.}          
\def\baas{BAAS}               

\def\icarus{Icarus}           
\def\jcap{J. Cosmology Astropart. Phys.}
\def\mnras{MNRAS}             
\def\pasp{PASP}               
\def\ssr{Space~Sci.~Rev.}     
\def\nat{Nature}              

\def\gca{Geochim.~Cosmochim.~Acta}   
\def\grl{Geophys.~Res.~Lett.} 
\def\jgr{J.~Geophys.~Res.}    

\def\planss{Planet Space Sci.}   
\def\procspie{Proc.~SPIE}   
\def\pnas{P. Natl. A. Sci.}   
\def\pasp{Publ. Astron. Soc. Pac.}   
\def\gca{Geochim Cosmochim Ac}       

\def\RMxAA{Rev. Mex. Ast. Astr.}    
\def\oleb{Origins Life Evol. B.}      
\def\epsl{Earth Planet Sc. Lett.}     
\def\asr{Adv. Space Res.}             
\def\ija{Int. J. Astrobiology}        
\def\jgrp{J. Geophys. Res-Planet}     
\def\jgra{J. Geophys. Res-Atmos}     
\def\aem{Appl. Environ. Microb.}      
\def\com{Curr. Opin. Microbiol.}       
\def\jm{J. Meteorol.}                   
\def\rg{Rev. Geophys.}                  
\def\crps{Crit. Rev. Plant Sci.}        
\def\am{Am. Mineral.}                   
\def\ptrslb{Philos. T. Roy. Soc. B.}    
\def\prslb{P. R. Soc. Lond. B. Bio.}    
\def\jas{J. Atmos. Sci.}                
\def\ag{Ann. Glaciol.}                  
\def\anac{Astron. Nachr.}               
\def\iau{Proc. Int. Astron. Union}      
\def\james{J. Adv. Model Earth Sy.}     
\def\areps{Annu. Rev. Earth Pl. Sc.}    
\def\fme{Fems. Microbiol. Ecol.}        
\def\mr{Microbiol. Rev.}                
\def\ea{Exp. Astron.}                   
\def\jp{J. Phycol.}                     
\def\pafs{Proc. Ann. French. Soc. Astron. Astrophy.}    
\def\grl{Geophys. Res. Lett.}           
\def\geo{Geol. Mag.}                    

\def\rearth{R$_{\oplus}$}

\begin{document}

\begin{frontmatter}



\title{The Habitability of Planets Orbiting M-dwarf Stars}

\author{Aomawa L. Shields$^{1,2,4}$, {Sarah Ballard}$^3$, {John Asher Johnson}$^4$}
\address{$^1$University of California, Irvine, Department of Physics and Astronomy, 4129 Frederick Reines Hall, Irvine, CA 92697, USA\\ 
$^2$University of California, Los Angeles, Department of Physics and Astronomy, Box 951547, Los Angeles, CA 90095, USA\\
$^3$Massachusetts Institute of Technology, 77 Massachusetts Ave, 37-241, Cambridge, MA 02139, USA\\
$^4$Harvard-Smithsonian Center for Astrophysics, 60 Garden Street, Cambridge, MA 02138, USA}

\begin{abstract}
The prospects for the habitability of M-dwarf planets have long been debated, due to key differences between the unique stellar and planetary environments around these low-mass stars, as compared to hotter, more luminous Sun-like stars. Over the past decade, significant progress has been made by both space- and ground-based observatories to measure the likelihood of small planets to orbit in the habitable zones of M-dwarf stars. We now know that most M dwarfs are hosts to closely-packed planetary systems characterized by a paucity of Jupiter-mass planets and the presence of multiple rocky planets, with roughly a third of these rocky M-dwarf planets orbiting within the habitable zone, where they have the potential to support liquid water on their surfaces. Theoretical studies have also quantified the effect on climate and habitability of the interaction between the spectral energy distribution of M-dwarf stars and the atmospheres and surfaces of their planets. These and other recent results fill in knowledge gaps that existed at the time of the previous overview papers published nearly a decade ago  by Tarter \emph{et al.} (\citeyear{Tarter2007}) and Scalo \emph{et al.} (\citeyear{Scalo2007}). In this review we provide a comprehensive picture of the current knowledge of M-dwarf planet occurrence and habitability based on work done in this area over the past decade, and summarize future directions planned in this quickly evolving field. 
\end{abstract}


\end{frontmatter}

\tableofcontents

\pagebreak
\section{Introduction}\label{sec1}
What was once the realm of science fiction\textemdash Earth-sized planets outside of the Solar System, where life might exist\textemdash is now scientific fact. At the time of writing, over three thousand confirmed planets have been discovered orbiting other stars\protect\footnotemark{}. Many of these planets are especially captivating because of their orbital distances, which place them in their stars' canonical habitable zone\textemdash the region around a star where an orbiting planet with an Earth-like atmosphere (CO$_2$-H$_2$O-N$_2$) could maintain water in liquid form on its surface \citep{Hart1979, Kasting1993}. The mass and radius distribution of these habitable-zone planets is power-law in shape, rising steeply toward smaller planets  with $R_p \lesssim 1.6$~\rearth\  \citep{Borucki2013, Quintana2014, Torres2015}. Thus, many of the small planets detected by NASA's \textit{Kepler} mission are likely to have rocky surfaces like the Earth \citep{Rogers2015}. On Earth, where there is water, there is life. Therefore such planets are exciting prospects to consider in the search for life outside of the Solar System. 

However, recent advances in theoretical research have revealed the myriad factors that contribute to determining a planet's habitability beyond orbital distance from its parent star. Specifically, the interaction between a star and orbiting planets can induce both radiative and gravitational effects on planetary climate \citep{Budyko1969, Barnes2008, Barnes2009, Barnes2013, Kopparapu2013a, Kopparapu2013b, Yang2013, Shields2013, Shields2014, Shields2016a}. These effects are now entering into the larger discussion of habitability. Additionally, an understanding of how these processes might change for different host stars and planetary system architectures has expanded and deepened, which will inform the prioritization of planets for follow-up by future characterization missions. Identifying the planetary systems that have the highest likelihood of hosting a water- and possibly life-bearing world, as well as the properties, whether stellar or planetary, that have the largest influence on long-term habitability, is of chief importance to the fields of astrobiology and exoplanet astronomy. \footnotetext{https://exoplanets.nasa.gov/ as of October 7, 2016}

Interest in M-dwarf stars as hosts for habitable planets has increased markedly over the last twenty years as the field of exoplanet discovery and characterization has grown. Since M dwarfs comprise $\sim$70\% of all stars in the galaxy \citep{Bochanski2010}, they offer the best chance of finding habitable planets through sheer numbers and proximity to the Sun. They also offer clear observational advantages. Small planets are easier to detect orbiting small stars via the radial velocity and transit techniques, as spectroscopic Doppler shifts and photometric transit depths are larger due to the smaller star-to-planet mass and size ratios, respectively. Also, because of the relatively low temperatures and luminosities of M dwarfs, their habitable zones are much closer to the stars than those of Sun-like stars, increasing the geometric probability of observing a transit \citep{Gould2003, Nutzman2008}, as well as the frequency of transits of habitable-zone planets during a given observational time period. Small planets orbiting small, low-mass stars are also better suited to the application of transmission spectroscopy methods to characterize their atmospheres (e.g. \citealp{Kreidberg2014}).  

Additionally, the lengthy stellar lifetimes of M-dwarf stars is a benefit. The ``dwarf" classification is assigned to stars that are in the main-sequence phase of stellar evolution, converting hydrogen to helium in their cores. Every M star that has ever formed is still on the Main Sequence. This is because M-dwarf stars, given their low masses, burn their nuclear fuel at vastly slower rates compared to Sun-like or brighter stars \citep{Iben1967, Tarter2007}. They are therefore extremely long-lived, with main-sequence lifetimes of trillions of years for the lowest-mass M dwarfs \citep{Laughlin1997}. They would therefore offer plentiful timescales for planetary and biological development and evolution on orbiting planets. 

However, the prospects for the habitability of planets orbiting M-dwarf stars have long been debated, due to key differences between the unique stellar and planetary environments around these stars, and the environments of hotter, brighter stars. Previous overview papers of nearly a decade ago summarized the current understanding of the instrumental requirements for observing M-dwarf planets \citep{Scalo2007}, and addressed outstanding questions concerning the impact on surface life of stellar flare activity, synchronous rotation, and the likelihood of photosynthesis on M-dwarf planets \citep{Tarter2007}. However, at the time of these papers, no terrestrial-sized planets had yet been discovered around M-dwarf stars, and no statistical information about the occurrence rate of exoplanets as a function of stellar type was available. Additionally, the use of multidimensional climate models to explore how planets orbiting stars other than the Sun achieve global energy balance as a function of their stellar and planetary environments was not prevalent.

Over the past decade, major progress has been made by both space- and ground-based observatories, and theoretical modelers, to more precisely measure the likelihood of small planets to orbit in the habitable zones of M-dwarf stars, and the effect on climate and habitability of the unique stellar and planetary environments of M dwarfs. About 200 exoplanets have been found around M-dwarf stars, many in their stars' habitable zones  \citep{Anglada2013, Quintana2014, Rowe2014, Torres2015, Crossfield2015, Barclay2015, Schlieder2016, Anglada-Escude2016}. These planets have been found through radial velocity (RV) methods, transit detection missions such as NASA's \emph{Kepler} mission \citep{Borucki2006} and the K2 mission\textemdash an ecliptic plane survey that re-purposes the \emph{Kepler} spacecraft following the loss of two of its reaction wheels \citep{Howell2014}\textemdash and gravitational microlensing \citep{Gaudi2012}. With the unveiling of the next generation of telescopes in the coming decade, we expect to find more of these planets, ushering in a new era that will use both observational and theoretical methods together to generate a prioritized target list of potentially habitable planets to follow up on with future missions to characterize their atmospheres. As discussed in the following sections, both observational data and modeling efforts indicate that the best place to look for habitable planets may be around M-dwarf stars \citep{Joshi1997, Joshi2003, Selsis2007, Wordsworth2010a, Joshi2012, Shields2013, Kopparapu2013c, Shields2014, Dressing2015, Gale2015}.

In this review we provide a comprehensive picture of the current knowledge of M-dwarf planet habitability, based on work done in this area over the past decade. We contextualize the specific, unique properties of the M dwarf spectral class (Section 2) before focusing primarily on the observational work that has been carried out to constrain the demographics of their planets (Section 3). We then provide an overview of the wide range of factors and processes that can influence planetary habitability (Section 4), and summarize recent theoretical work done to quantify their effect on the habitability of M-dwarf planets in particular (Section 5). Finally, we discuss possible niches for life on M-dwarf planets given the mechanisms employed by life to survive within the types of environmental extremes on the Earth that may be possible on M-dwarf planets (Section 6). We leave an in-depth discussion of planned efforts to detect atmospheric biosignatures with upcoming missions for future review papers.

\section{Stellar Astrophysics: The M-dwarf Spectral Class}\label{sec2}
To first order, stars of a given age are a two-parameter family. Mass and chemical composition inform not only the formation, evolution, and fate of stars, but the corresponding narratives for whether and how planets form around them (e.g. \citep{Johnson10}). While the Solar System has long furnished our default blueprint for planet occurrence in the Milky Way, exoplanets populate M dwarfs very differently than Sun-like stars. 

M dwarfs are our galaxy's silent majority: they constitute 70\% of the stars in the Milky Way \citep{Reid97,Bochanski2010} and 40\% of its stellar mass budget \citep{Chabrier03}, yet not a single M dwarf is visible to the naked eye. They span nearly an order of magnitude in mass and two orders of magnitude in luminosity, bracketed by the hydrogen burning limit of 0.08 $M_{\oplus}$ \citep{Baraffe96} on the low-mass end and extending to half the mass of the Sun (see Section 3.1 for a discussion of the M spectral type and links to physical properties). The transition within stars from partially to fully convective ($<$0.35 $M_{\odot}$) also occurs midway through the M spectral type \citep{Chabrier97}. As a spectral class, M dwarfs span a larger range in mass than the next three spectral classes (FGK) {\it combined} \citep{Boyajian12,Boyajian13}. 

\subsection{Defining the M spectral type}
Spectral type M is defined by the presence of strong absorption features due to the diatomic molecule titanium oxide (TiO) at blue and green optical wavelengths ($\sim$4500--5700 \AA) \citep{Morgan43}. In fact, M dwarfs are cool enough that their spectra are heavily veiled throughout by molecular bands not only of TiO, but also of diatomic hydrogen (H$_{2}$), water (H$_{2}$O), and vanadium oxide (VO) to the extent that the true continuum is obscured. As with other classification schema, the boundaries between spectral classes reflect the particulars of astronomical history. Strong absorption of TiO also appears at redder wavelengths in stars ``earlier" (corresponding to hotter stars) than M0, but that knowledge awaited red-sensitive detectors \citep{Kirkpatrick91, Bessel91}. The M spectral sequence is bracketed at the low-mass end at M7V/M8V, corresponding to the smallest star capable of fusing hydrogen into helium (0.08 $M_{\odot}$) \citep{Chabrier97}. However, the very bottom of the Main Sequence is challenging to link exactly to spectral type. Small mass or metallicity perturbations astride the star/brown dwarf border can result in physical changes large enough to move objects from one category to another \citep{Dieterich14}. Spectral types as late as L2V \citep{Dieterich14} may sometimes possess the minimum mass necessary for hydrogen fusion to occur. 

The spectral slope and the strengths of molecular and atomic features differentiate M dwarf spectral types from one another, though metallicity variation makes this a subtle endeavor \citep{Gizis97}.  A full suite of spectral ``standard" stars from 6300 to 9000 \AA ~enabled the technique of least-squares comparison for spectral typing \citep{Kirkpatrick91}. Comparison between the spectral indices of the target and those measured from benchmark spectral types is now the {\it de facto} industry standard, though several methodologies exist \citep{Reid95, Hawley1996, Covey07}. We describe the challenge of modeling M dwarf spectra in greater detail in Section \ref{sec:mdwarf_spectra}. The painstaking and decades-long process of anchoring spectral type in physical properties like mass and radius is described in more detail in Section \ref{sec:mdwarf_mr}.


\subsection{M dwarf evolution and lifetimes}
M dwarfs have rightfully been called ``models of persistence" \citep{Laughlin1997}. A scaling argument demonstrates why this title is apt: if a star's main-sequence lifetime scales as its mass divided by its luminosity $M_{\star}/L_{\star}$, with $L_{\star}\sim M_{\star}^{3}$ at the bottom of the Main Sequence \citep{Prialnik09}, it stands to reason that a 0.1 $M_\odot$ star lifetime should be at least 100 times that of our Sun. In fact, a 0.1 $M_\odot$ star will burn hydrogen into helium for 12 trillion years \citep{Laughlin1997}, roughly $10^3$ solar lifetimes, because of an additional source of longevity among small stars. While a typical Sun-like star will use only 10\% of its nuclear fuel, the fully convective interiors of M dwarfs with masses below 0.35 $M_{\odot}$ can dredge up and burn nearly all of their reserves \citep{Adams04}. M dwarfs at the lowest end of the Main Sequence, $<0.16$ $M_{\odot}$, will never become red giants \citep{Laughlin1997}. Their long lifetimes, not to mention their ubiquity, favor M dwarfs at least statistically for being the likeliest sites for the evolution of life (we revisit this idea in Section 5) \citep{Loeb16}.

The signatures of magnetic activity in the atmospheres of M dwarfs are linked to their age and physical properties. Decades of research have elucidated these links, anchored to observables such as X-ray emission, H$\alpha$ in emission and other spectral features, rotation rates, and spot coverage \citep{Giampapa86, Soderblom91, Reid95, Hawley96, Delfosse98, Hawley99, Barnes2004, Oneal04,  Irwin11}. All of these terms quantify stellar ``activity," a term used to describe phenomena associated with the strength of the star's magnetic field. 

The rotation of M dwarfs, both fully and partly convective, is dependent upon the age and mass of the star. A trade-off between physical processes early in the lives of M dwarfs must occur to match the rotation rates observed in young clusters: Kelvin-Helmholtz contraction and late-stage accretion spin the star up, while angular momentum is lost from interaction with the disk \citep{Hartmann89, Bouvier97, Koenigl91, Colliercameron95, Matt05}. Subsequent spin-down after reaching the Main Sequence is dominated by magnetized winds \citep{Barnes03}. There is evidence among Sun-like stars that rotation rate can be used as a clock: because spin-down due to magnetic braking occurs in a predictable fashion, a star's present-day rotation (among other activity indicators) encodes its age \citep{Stauffer87, Barnes96, Krishnamurthi97, Mamajek08}. For FGK dwarfs older than 500 Myr, rotation period decreases as $t^{-1/2}$ \citep{Barnes03, Barnes07}. However, there is additional subtlety to the science of ``gyrochronology" for M dwarfs. M dwarfs of a given mass (provided it is above the convective limit) exhibit a coherent spin-down \citep{McQuillan13, McQuillan14}. However, below the convective limit there appear two divergent populations of faster and slower rotators \citep{Irwin11, Newton16b}. This gap is likely attributable to a period of short but rapid spin-down of mid-to-late M dwarfs. While they maintain rapid rotation for several gigayears, they reach periods of 100 days or more by a typical age of 5 Gyr \citep{Newton16b}.

Other observable signatures of activity mark the evolution of M dwarfs. The presence of H$\alpha$ in emission (often coincident with X-rays in emission \citep{Reid95, Covey07}) signals the presence and strength of magnetic activity and decreases with age \citep{West06, West2008}. Across the M spectral class, the ``active" duration of a star's life varies from 1 Gyr in the case of M0 dwarfs to 8 Gyr or more for spectral type M8 \citep{West06}. We discuss the relationship between activity and habitability in Section 5.1.

\subsection{Complicated spectral properties}
\label{sec:mdwarf_spectra}
Inferring the physical properties of M dwarfs from spectra, upon which a measurement of planetary radius and equilibrium temperature hinges so critically, presents difficulties on multiple fronts. The direct comparison of theoretical spectra to observations, robust for deducing the properties of solar-type stars, is challenging for low-mass stars. Such spectra rely on detailed, computationally intensive modeling of convection in low-mass stellar interiors \citep{Mullan01, Browning08} and complete lists of the complex array of molecules and grains that reside in their atmospheres \citep{Tsuji96, Allard00}. The presence of molecules such as H$_{2}$, H$_{2}$O, TiO, and VO present major complications for radiative transfer calculations, both because their transitions are numerous and their absorption coefficients frequency-dependent (for more massive stars, gray approximations are standard practice) \citep{Chabrier97}. The high pressure in M dwarf atmospheres means that collision-induced transitions must additionally be considered \citep{Chabrier97}. For these reasons, critical inroads linking spectral features to physical properties of M dwarfs have often resulted from empirical study \citep{Torres13}. This challenge is compounded by the possibility that stellar properties may also depend on other parameters in a significant way. Stellar activity and metallicity are two such parameters: we define the first in Section 3.2, and the second encodes the fraction of mass of the star in the form of elements heavier than hydrogen or helium. 

M dwarfs in wide binaries with FGK dwarfs, whose metallicities can be measured via comparison of observed spectra to theoretical models, provide crucial benchmarks for measuring M dwarf metallicity. Bonfils et al. (2005) innovated a metric for retrieving M dwarf metallicities from broadband photometry with a subset of 20 such binaries \citep{Bonfils05}. They assumed that both members of the binary formed from the same molecular cloud and therefore possess the same fraction of metals. They then identified the empirical color-magnitude relationship among the M dwarfs that successfully traced the metallicities of the more massive companions, although subsequent studies refined this relationship and placed it on an absolute scale \citep{Johnson09, Schlaufman10, Neves14}. Rojas-Ayala et al. (\citeyear{Rojasayala12}), with a similar motivation, identified an empirical relationship for both metallicity and temperature with $K$ band spectral features (or combinations thereof) in M dwarfs. This technique has since been extended into $J$ band, $H$ band, and optical wavelengths \citep{Terrien12, Onehag12, Mann13, Neves14}. These metrics rely upon spectral indices: the ratio of fluxes from different parts of the spectrum, as well as equivalent widths, which are measures of the strength of {\it individual} molecular or atomic absorption features (specifically, the $\Delta \lambda$ centered on the $\lambda_{0}$ for the transition within which a rectangle encloses the same area as the absorption line). Newton et al. (2015) have since demonstrated that equivalent widths alone of some M dwarf spectral features encode not only metallicity, but their stellar radii and effective temperature as well \citep{Newton15}. Recent work, which we describe more in Section 3.1, suggests that metallicity metrics contain a hidden dependence upon carbon-to-oxygen ratio \citep{Veyette16}. 

\subsection{Mass-Radius Relationship}
\label{sec:mdwarf_mr}
Theoretical modeling of M dwarfs is tasked not only with the creation of realistic synthetic spectra (described in the previous section), but also the accurate reproduction of their masses and radii. These properties depend on the interior structure of stars.  Here too, we rely upon empirical benchmarks. M dwarfs in eclipsing geometries with other stars directly furnish mass, radius, and effective temperature (e.g. \citep{Andersen91, Torres13, Morales09, Carter11}). For M dwarfs near enough to have their diameters directly measured via interferometry, we have direct access to radius and temperature \citep{Lane01, Segransan03, Berger2006, Boyajian12,vanbelle09, vonbraun11, vonbraun12}. 


There exists a tension between stellar models and empirically-derived properties for low-mass stars at the level of 5-10\% \citep{Torres02, LopezMorales07, Morales08, Morales09,Vida09, Carter11, Kraus11}. These models typically underpredict the radii and overpredict their temperatures. This result is robust to other possible observational biases such as tidal inflation in binaries, though the effect is smaller when the binary is widely separated \citep{Kraus11, Irwin11}. 

Viewed through the lens of exoplanetary science, precise knowledge of the planetary properties hinges on knowledge of the mass-radius relation for M dwarfs. A discrepancy of 10\% near the 1.5 $R_{\oplus}$ value, for example, profoundly affects the interpretation of the planet's composition \citep{Rogers2015}. In the case of the first Earth-sized planets observed to transit an M dwarf, the authors employed a proxy star of the same spectral type with directly-measured properties \citep{Muirhead12}. Ultimately, {\it distance} measurements to nearby M dwarfs will reduce the systematic error on their radii and refine their uncertainties (and correspondingly, the radii of their planets). Dittmann et al. (2014), using distances measured with parallax for MEarth M dwarfs, found they were on average 0.08 $R_{\odot}$ larger than the radius estimates derived with no distance information \citep{Dittmann14}. Forthcoming parallaxes measured with ESA's {\it Gaia} mission \citep{Lindegren10} will furnish radius measurements of the 4$\times10^{5}$ M dwarfs within 100 pc to within 2-3\% \citep{Sozzetti14,Sozzetti15}.  

\subsection{Planet Detectability}
\label{sec:detect}

Several factors conspire to make M dwarfs the predominate sites for exoplanetary study in the next decade. This is particularly true for exoplanets residing in the stars' habitable zones (we consider the definition of this zone in detail in Section \ref{sec4}). Planets orbiting M dwarfs produce a larger observational signature. The same size planet will induce a larger reflex motion on a lower-mass host star than a Sun-like star (2.5 times larger for a 0.25 $M_{\odot}$ M4V dwarf), and block a larger fraction of starlight in the transit geometry (1.3 mmag for an Earth transiting an M4V dwarf, versus 0.084 mmag for the Sun) \citep{Charbonneau07}. But crucially, the smaller luminosity of M dwarfs means that potentially habitable planets can reside much closer to the host star. The transit probability of planets in the habitable zone of M dwarfs is corresponding higher as well. The probability of a transit for a planet residing in the habitable zone is 1.5\% (M4V dwarf) and 2.7\% (M8V dwarf), significantly above the Earth-Sun value of 0.47\% \citep{Charbonneau07}.

However, though planets orbiting M dwarfs produce larger observational signatures than the same planet orbiting a larger star, M dwarfs present a unique set of challenges for planet detection. Stellar magnetic activity can confound the interpretation of M dwarf radial velocities. Activity in the form of both starspots and convective inhomogeneities can mimic exoplanetary signatures at the rotational period of the star or its harmonics. When star spots rotate onto the limb of the star, they can change the spectral line profiles and therefore its centroid \citep{Saar97}. Localized magnetic fields associated with starspots can also modify convective flow, producing a net blueshift \citep{Gray09, Meunier10}. While stellar activity presents challenges to radial velocity detection of planets around stars across the Main Sequence, only for M dwarfs does the characteristic rotation period coincide with the habitable zone. Newton et al. (\citeyear{Newton16}) addressed this specific concern in their census of the rotation periods of early-to-mid M dwarfs \citep{Newton16}. In this work and similar work by Vanderburg et al. (2016), they found that typical rotation periods of M dwarfs with spectral type earlier than M4V overlap with the orbital periods of planets in their habitable zones (see Figure 1) \citep{Vanderburg16}.

\begin{figure}
\begin{center}
\includegraphics [scale=1.00]{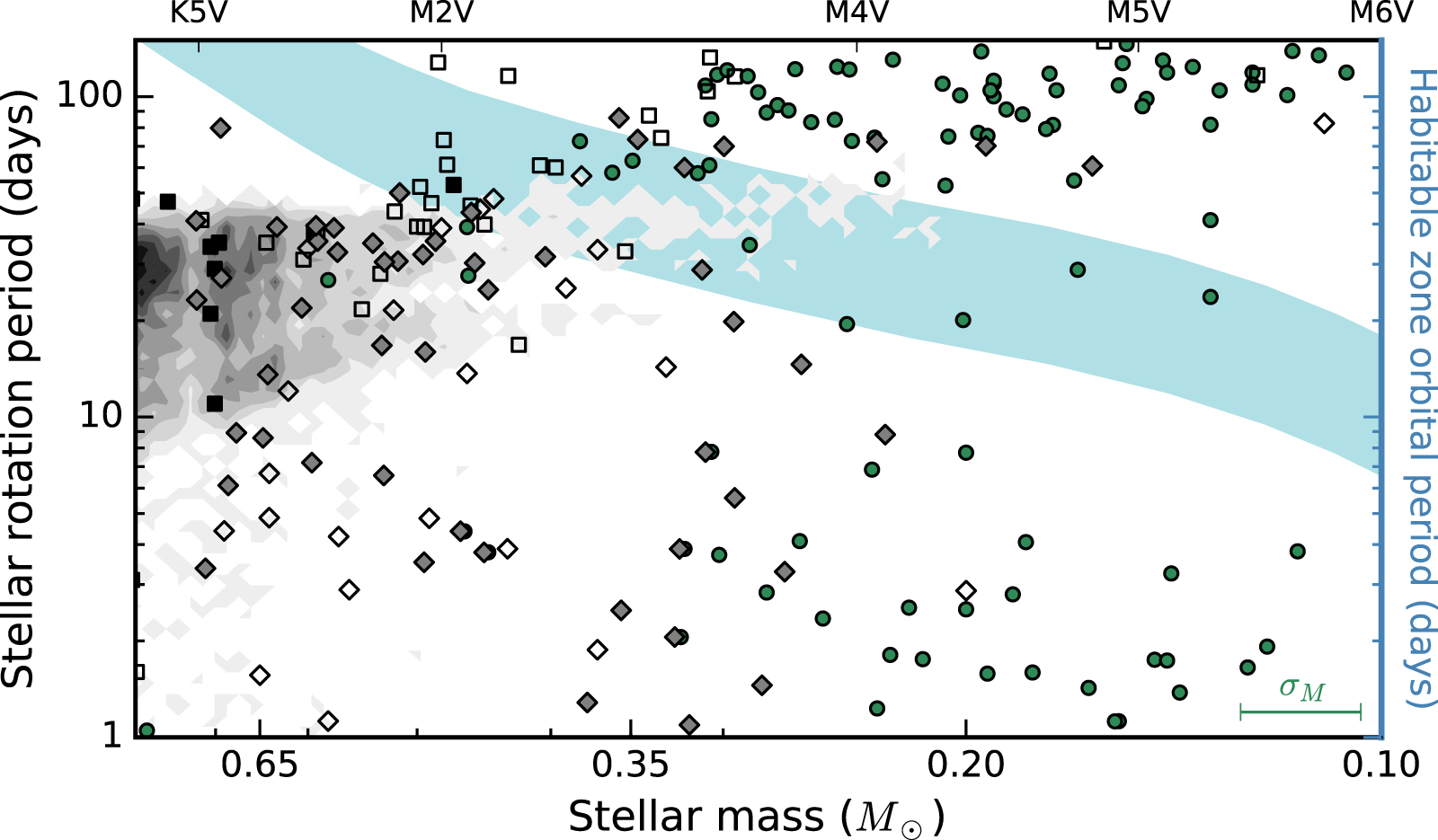}\\
\caption{Figure 1 from Newton \textit{et al.} (\citeyear{Newton16}), showing stellar rotation period versus mass for late K and M dwarfs. The divergence of stars into an upper and lower envelope with rotation period is clear below the fully-convective mass of 0.35 $M_{\odot}$. The blue shaded region depicts the range of orbital periods corresponding for the stellar habitable zone. Possibilities for confusion of activity and planetary signals are highest where their periodicities overlap. Mid-to-late M dwarfs, which rotate either faster or slower than a habitable planet orbits, are at a distinct advantage. This figure is reproduced from \citet{Newton16} with permission from the authors and AAS. } 
\label{Figure 1. }
\end{center}
\end{figure}

M dwarf planet detection with the transit method, in contrast, is less impeded by stellar variability. Stellar flares in light curves can potentially complicate the search for transits, and have even been observed during transit events themselves \citep{Kundurthy11}. However, studies of M dwarf photometric variability on the timescales of transits conclude that stellar activity should not meaningfully affect the detection of exoplanets \citep{Tofflemire12, Goulding12} (though they caution that this depends upon photometric filter and spectral type).

These challenges notwithstanding, we know now of hundreds of planets orbiting M dwarfs. We describe our current understanding of planet occurrence orbiting M dwarfs in the next section.

\section{Observational Landscape and Demographics of M-dwarf Planets}\label{sec3}
\subsection{RV Planets and the dearth of Gas Giants, metallicity effects}

Hot and warm Jupiters orbiting Sun-like stars constituted the bulk of planet discoveries for 20 years. Stars with masses close to the Sun are well-suited to radial velocity searches: they are bright, rotate slowly, and their spectra contain a forest of narrow absorption lines well-suited for measuring high-precision Doppler shifts at optical wavelengths \citep{Johnson10}. Yet, within 3 years of the very first radial velocity detection of an exoplanet, the first planet orbiting an M dwarf was uncovered around GJ 876 \citep{Delfosse98, Marcy98}. It was only the ninth planet to be discovered around a Main Sequence star \citep{Bonfils07}. Before M dwarfs were revealed to host small planets ($<4$ $R_{\oplus}$) in abundance (which we describe in detail in Section \ref{sec:occur}), their dearth of gas giants established them as comparatively poor planet hosts \citep{Laws03, Johnson07, Lovis07, Johnson10}.  To date there is still only a single known hot Jupiter orbiting an M dwarf \citep{Johnson12}, so the earliest era of exoplanetology was heavily focused upon larger stars. 

The sample of planets detected with radial velocity around M dwarfs is comparatively small, yet rich in science. The first occurrence rate calculations for M dwarfs with RV used only 11 planet detections \citep{Bonfils13}. In comparison, later that same year, the occurrence rate from transits employed more than 100 detections from NASA's {\it Kepler} Mission \citep{Dressing2013}. In fact, both of these measurements postdate the first planet occurrence rate for M dwarfs, which was in fact calculated from microlensing detections (described in more detail in Section 3.3) \citep{Cassan12}.  The finding that metal-rich stars are likelier to host Jovian planets \citep{Fischer05} at first hinted that a metallicity offset between Sun-like stars and M dwarfs could explain the discrepancy \citep{Bonfils05}. However, though stars in the solar neighborhood possess similar metallicities, M dwarfs still host 2-4 times fewer Jovian planets \citep{Johnson07}. 

As radial velocity sensitivity to M dwarfs improved, both with use of larger telescopes \citep{Butler04, Bonfils05}, infrared spectroscopy \citep{Bean09}, and longer time baselines, their population of smaller planets came into focus. The landmark discovery of the 1.3 $M_{\oplus}$ temperate planet orbiting the closest star to Earth resulted from a 16-year-long radial velocity campaign of 0.12 $M_{\odot}$ M dwarf Proxima Centauri \citep{Anglada-Escude2016}. Even with the 1 m s$^{-1}$ precision possible with the HARPS spectrograph at La Silla over many years \citep{Pepe11}, the iron-clad discovery of Proxima b hinged upon nearly nightly observations of the star from 2016 January--March \citep{Anglada2013}. Its discovery highlights yet another challenge to radial velocity searches. While the semi-amplitude of Proxima b at 1.4 m s$^{-1}$ is not especially small \citep{Pepe11}, it eluded detection for years because of the sparse radial velocity time sampling coupled with the long stellar activity cycle \citep{Anglada-Escude2016}. The procedure for detangling aliases from orbital frequencies is very sensitive to the time sampling of observations \citep{Dawson10}. This is particularly salient for M dwarfs, whose typical rotation periods can overlap with the orbital periods of habitable-zone planets (see Section 2.5). 

Between 1--4 $R_{\oplus}$, the planet-metallicity correlation is weaker across spectral type \citep{Sousa08, Mayor11, Schlaufman11}. Some studies find that a typical protoplanetary disk produces a continuum of planets in this range, with less to no effect from host star metal content \citep{Schlaufman15}. Others find that stellar metallicity does play a role for planets this size, particularly with respect to whether more Earth-like or Neptune-like planets result \citep{Buchhave2014, Wang15}. One major caveat in the endeavor to link stellar metallicity to planet occurrence is detection method bias. The preferential sensitivities of the radial velocity and transit techniques to different types of planets (i.e., dense low-mass planets versus gaseous sub-Neptunes, respectively \citep{Wolfgang12}) may also confound a simple interpretation. 

More recent work has examined the relationship between metallicity and  carbon-to-oxygen ratio as inferred from spectra. Carbon and oxygen abundances affect the pseudo-continua near spectral metallicity indicators, so that C/O enrichment can signficantly bias metallicity measurements \citep{Veyette16}. The empirical planet-metallicity correlation  may therefore also encode a dependence upon C/O ratio. There exists a strong theoretical framework for such an effect, based upon the hypothesis that planetesimal chemistry (and indeed, the presence or absence of planets) reflects the composition of oxygen-rich dust inherited from the molecular cloud \citep{Barshay76, Bond10, Moriarty14, Gaidos2016}.  

\subsection{MEarth, GJ 1214b, and TRAPPIST}
Up until 2009, the landscape of transiting exoplanets was confined to a single planet smaller than Neptune. CoRoT-7b, orbiting a G dwarf 80\% the mass of the Sun, was discovered with ESA's CoRoT Mission and was the first rocky planet outside of the Solar System to have a measured density \citep{Leger09, Queloz09}. So-called super-Earths, intermediate in size between the Earth and Neptune, eluded wide-field transit surveys of Sun-like stars. Their typical precision over a transit duration was several times larger than the transit of a rocky planet (0.001 mag or 1000 ppm precision, in contrast with the 300 ppm transit of CoRoT-7b itself). 

The MEarth Observatory was designed to detect planets as small as 2 $R_\oplus$ from the ground with off-the-shelf components, simply by targeting M dwarfs rather than Sun-like stars \citep{Charbonneau07, Nutzman08}. A potentially rocky planet sihouetted against an M dwarf can produce a transit readily detectable from the ground, where its counterpart around a Sun-like star might be detectable only from space (see Section \ref{sec:detect} for additional discussion about the detectability of planets orbiting M dwarfs). There existed only one planet known to transit an M dwarf before MEarth:   GJ 436b was first detected with radial velocity measurements \citep{Butler04,Maness07} and only discovered to transit afterward \citep{Gillon07}. The high proper motion of the nearby M dwarfs on MEarth's curated target list offered an additional advantage. The high false-positive rate challenging ground-based transit surveys is due to blended eclipsing binaries. When combined with the light of the target star, these binaries can produce stellar eclipses shallow enough to masquerade as planets. MEarth's stars, however, possess high enough proper motions that a blend could be {\it a priori} ruled out by examining the field before or afterward for evidence of additional stars at the location of the planet host.

The discovery of GJ 1214b was the first of its kind, a transiting Super-Earth detected from the ground, and a super-Earth transiting an M dwarf \citep{Charbonneau09}. The brightness and size of the host star suddenly placed atmospheric studies within reach for an exoplanet smaller than Neptune. The first transmission spectrum for GJ 1214b was gathered from the ground with the Very Large Telescope facility \citep{Bean2010}, and hundreds of hours of observations from NASA's {\it Hubble} and {\it Spitzer} missions followed \citep{Berta12, Desert11, Kreidberg2014, Fraine13}. The flatness of the transmission spectrum resists conclusive interpretation, but the first transiting super-Earth to orbit an M dwarf is likely a cloudy or hazy world \citep{Kreidberg2014}. 

The TRAPPIST survey has already uncovered three small planets by similarly monitoring the brightnesses of small stars \citep{Gillon16}. The transits of these rocky planets appeared after 2 years of observation with TRAPPIST, a 60 cm prototype telescope that will ultimately comprise part of the SPECULOOS observatory (see Section 3.6). While MEarth focuses upon mid-M dwarfs, TRAPPIST targets spectral types later than M6V (cooler than 3000 K \citep{Boyajian12}). The very existence of the three planets orbiting the 0.12 $R_{\odot}$ star TRAPPIST-1 affirms that planet formation proceeds all the way to the bottom of the Main Sequence \citep{Gillon16}. 

\subsection{Microlensing and Direct Imaging}
While the radial velocity and the transit detection methods favor close-in planets, both microlensing and direct imaging are best suited to planets at wide separations. There exists some overlap in detectability for radial velocity and microlensing searches, but the methods are mostly disjoint \citep{Clanton14b}. The typical M dwarf planet detected via microlensing is Saturn-sized and orbits 2.5 AU from its 0.5 $M_{\odot}$ host star \citep{Clanton14a}. Such a star would present a 5 m s$^{-1}$ peak-to-peak reflex motion over 7 years at the median inclination, challenging but detectable for radial velocity surveys like the California Planet Survey (CPS) and for HARPS \citep{Clanton14b, Johnson10, Bonfils13}. However, the planetary mass function inferred from microlensing surveys declines steeply for smaller planets \citep{Sumi10}. Marginalizing over the period and mass parameter space of planets plumbed by microlensing surveys, the posterior distribution in semi-amplitude peaks at 0.2 m s$^{-1}$. Thus, the scarcity of giant planets orbiting M dwarfs inferred from RV searches alone is due to their small signatures with that detection method. In fact, M dwarfs host on average of 0.15$^{+0.06}_{-0.07}$ giant planets per star (out to periods of 10$^{4}$ days) when microlensing detections are included; radial velocity results alone would place that occurrence rate closer to 3\% \citep{Clanton16, Bonfils13, Johnson10, Johnson10b}. However, with all detection methods accounted for, the occurrence rate of giant planets orbiting M dwarfs is still 4 times smaller than for FGK dwarfs \citep{Clanton16}. 

Direct imaging presents another set of constraints on the occurrence of M-dwarf planets Jupiter-sized and larger. Current calculations place the occurrence rate at 6.5\%$\pm$3.0\% for planets between  1 $<$ $m/M_{J}$ $<$ out to 20 AU (orbital periods of $\approx10^{2}$ years) \citep{Montet14}, and there exists an upper limit of 10\% when considering orbits out to 100 AU \citep{Bowler15}. Fewer than 6.0\% of M dwarfs host massive gas giants in the 5-13 $M_{\mbox{Jup}}$ range \citep{Bowler15}. 

\subsection{The Kepler Sample}
NASA's {\it Kepler} Mission was launched in May 2009 into an Earth-trailing orbit, where it would spend four years dutifully monitoring thousands of stars for transit signals \citep{Borucki2006}. Of the approximately half a million stars brighter than 16th magnitude and residing in {\it Kepler}'s field of view, telemetry concerns limited the {\it Kepler} census to only a fraction of these. The selection process from {\it Kepler}'s Input Catalog \citep{Brown11} reflected its science driver to establish the frequency of Earth-like planets orbiting Sun-like stars: 90,000 of the 150,000 stars in the {\it Kepler} sample were G type stars on or near the Main Sequence \citep{Batalha10}. The calculus of star selection also favored detectability of transits in {\it Kepler}'s optical bandpass. While the larger transits of M dwarfs made these appealing under that criterion, the feasibility of radial velocity follow-up strongly favored optically bright stars \citep{Batalha10}. The balance between these factors resulted in 3000 M dwarfs joining the {\it Kepler} sample, heavily weighted toward early spectral types \citep{Dressing2013}. We discuss the yield of M-dwarf planets uncovered by {\it Kepler} in the next section.

\subsection{Occurrence Rates}
\label{sec:occur}
The dearth of close-in gas giants orbiting M dwarfs hints at planet formation proceeding differently around smaller stars.  Indeed, both the character of an average planet and an average {\it system} of planets differ between M dwarfs and Sun-like stars. 

The sensitivity of NASA's {\it Kepler} Mission to small planets revealed the stark difference in planetary budget for M dwarfs as compared to hotter stars. The fact of M dwarfs hosting fewer giant planets than Sun-like stars was clear prior to {\it Kepler}'s launch, but {\it Kepler} uncovered that they host many more {\it small} planets. Early results for orbital periods interior to 50 days showed that planets between 2 and 4 $R_{\oplus}$ are twice as numerous around M dwarfs than Sun-like stars \citep{Howard12}. In fact, data from the first year and a half of Kepler's nominal 4 year mission showed that M dwarfs host an average of 0.90$^{+0.04}_{-0.03}$ planets per star between 0.5 and 4 $R_{\oplus}$ \citep{Dressing2013}. That finding has since been extended to planets out to an AU \citep{Dressing2015, Morton14, Mulders2015c, Gaidos16}, within which orbital period a typical M dwarfs hosts 2.0$\pm$0.45 planets per star \citep{Morton14} (consistent with the microlensing result of 1.9$\pm0.5$ planets per M dwarf between 1 and 10$^{4}$ $M_{\oplus}$ \citep{Clanton16}). There exist 3.5 times more small planets (1.0--2.8 $R_{\oplus}$) orbiting M dwarfs than main-sequence FGK stars, but two times fewer Neptune-sized ($>2.8$ $R_{\oplus}$) and larger planets \citep{Mulders2015c} (see Figure 2). Given the ubiquity of M dwarfs in the Milky Way, this finding has profound implications for the proximity of the nearest small exoplanet to Earth. Results even with a subset of the earliest {\it Kepler} data found that the nearest transiting Earth-size planet in the habitable zone of a cool star must lie within 21 pc of the Sun \citep{Dressing2013}. With the entirety of the 4 year nominal mission data in hand, that number was revised down to 10 pc \citep{Dressing2015}.

\linespread{1.15}
\begin{figure}
\begin{center}
\includegraphics [scale=0.60]{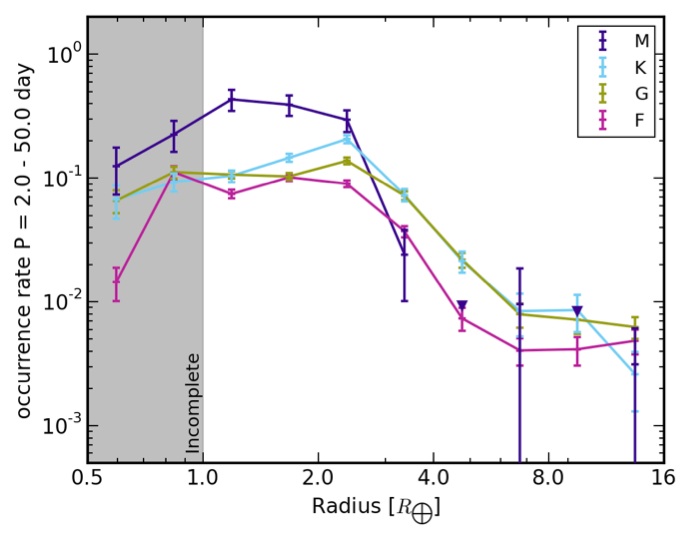}
\includegraphics [scale=0.60]{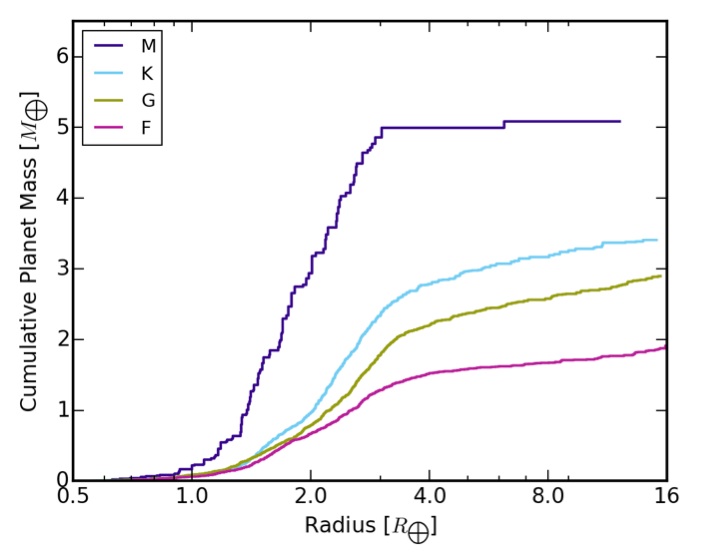}\\
\caption{Figure 7 from Mulders \textit{et al.} (\citeyear{Mulders2015c}), showing planet radius distribution vs. planet occurrence (top) and cumulative planet mass per star (bottom) for orbital periods between 2 and 50 days and M-, K-, G-, and F-dwarf host stars. This figure is reproduced from \citet{Mulders2015c} with permission from the authors and AAS.} 
\label{Figure 2. }
\end{center}
\end{figure}

Occurrence rates define a ``mean" planetary system. But in fact, no one model for a system of planets (characterized by a typical number of planets, with orbital inclinations drawn from a typical Rayleigh distribution) adequately recovers {\it Kepler}'s ensemble statistics. Stars hosting more than one transiting planet offer critical insight here. They allow for study not only of planet occurrence, but also for system architectures: transit duration and period ratios probe the mutual inclinations and spacings of adjacent planets. For Kepler's multi-planet systems, flat (mutual inclinations drawn from a Rayleigh distribution with $\sigma<3^{\circ}$) and manifold models are apt descriptors \citep{Lissauer11,Tremaine12,Fang12,Fabrycky14,Swift13}. However,the best--fitting models to the Kepler yield underpredict the number of singly--transiting systems by a factor of two \citep{Lissauer11,Hansen13}. This feature of the {\it Kepler} multi-planet ensemble is termed ``the {\it Kepler} dichotomy."

For M dwarfs in particular, the flat and manifold ($N\ge5$ planets mutually inclined by an average of 2$^{\circ}$) model occurs in 45$^{+12}_{-23}$\% of planetary systems, while the other planetary systems host either a single planet, or multiple planets with average mutual inclinations $>10^{\circ}$ \citep{Ballard16}. In contrast, for Sun-like stars, only 24\%$\pm7$\% of systems are arranged in this dynamically cool fashion \citep{Moriarty15}.

\subsection{Future Observational Work: Gaia, TESS and Beyond}

NASA's Transiting Exoplanet Survey Satellite (TESS), to be launched in late 2017, will unite the advantages of wide-field transit surveys with the precision and duty cycle of photometry from space \citep{Ricker2009}. The yield of small planets will be rich: of the 1700 transiting planets TESS is expected to uncover, 500 will possess radii smaller than 2 $R_{\oplus}$. A typical TESS target star receives 27 days of continuous observation, so the sensitivity of the mission strongly favors short periods \citep{Sullivan15}. A handful of transits of a small planet will be detectable over this duration only if those transits are individually large, which is why 75\% of small planets detected by TESS are expected to orbit M dwarfs \citep{Sullivan15}. It is likely that {\it every} small planet detected by TESS to reside in its star's habitable zone will orbit an M dwarf \citep{Sullivan15}, making M dwarfs almost certainly the sites for focused follow-up atmospheric study with the James Webb Space Telescope (JWST). 

ESA's CHEOPS mission \citep{Broeg13} and potentially ExoplanetSAT \citep{Smith10} will also monitor nearby stars for transit signatures in short order from space, while ExTrA \citep{Bonfils14} will combine spectroscopy with photometry to improve their ground-based transit search. The SPECULOOS observatory will monitor M dwarfs of spectral type M6 or later for transits \citep{Gillon13}. ESA's PLATO mission, to be launched in the 2022-2024 timeframe, will be sensitive to transits of M-dwarf planets even beyond the snow line \citep{Rauer14} . 
ESA's Gaia Mission, launched in 2013, will have a manifold impact upon M dwarf and exoplanetary study. Using the astrometric detection method, the mission is sensitive to Saturn mass planets around M dwarfs out to 25 pc that reside between 1 and 4 AU of their host stars \citep{Sozzetti14}. Moreover, the precision determination of the radii and distances of {\it known} M dwarf planet hosts to 5\% \citep{Bailerjones05} will refine in turn our precision of the sizes of their planets.  

A suite of spectrographs will also train their gazes on low-mass stars. MINERVA-Red, a photometric and spectroscopic observatory at Mt. Hopkins, will ultimately conduct its M dwarf planet search nearly autonomously \citep{Blake15}. The SPIRou spectropolarimeter (SpectroPolarim\'etre Infra-Rouge, R$\sim~$75,000) to be mounted on the 3.6m Canada France Hawaii Telescope (CFHT) will search for radial velocity signatures of planets orbiting low mass stars \citep{Santerne13}. The detection of the magnetic fields of the stars themselves with SPIRou will enable the detailed study of the impact of these fields upon planet occurrence \citep{Artigau14}. CARMENES, an \'echelle spectrograph (R$\sim$82,000) designed for the 3.5 m Calar Alto Telescope, will specifically target 300 nearby M dwarfs later than M4 \citep{Quirrenbach14}, similar to the target strategy of the Habitable Planet Finder on the 10 m Hobby-Eberly Telescope \citep{Mahadevan10}.

\section{Habitable Planets: Beyond Equilibrium Temperature}\label{sec4}
The primary step in classifying a planet as ``potentially habitable" has historically been to identify a planet that orbits at a particular distance from its host star to place it in the star's habitable zone. This is because for extrasolar planets the presence of surface liquid water is considered to be the most important indicator of a habitable planet, as all life on Earth uses it as a solvent for chemical reactions \citep{DesMarais2008, Cockell2016}. The habitable zone for stars of different masses is shown in Figure 3. While ice-covered planets could harbor life in the ocean beneath a frozen surface \citep{Tajika2008}, it would be challenging to detect the presence of sub-surface life remotely. Planets with conditions amenable to life on the surface offer the best chance for the spectroscopic detection of gases expelled into the atmosphere that are uniquely biological in origin.  

\linespread{1.15}
\begin{figure}
\begin{center}
\includegraphics [scale=0.45]{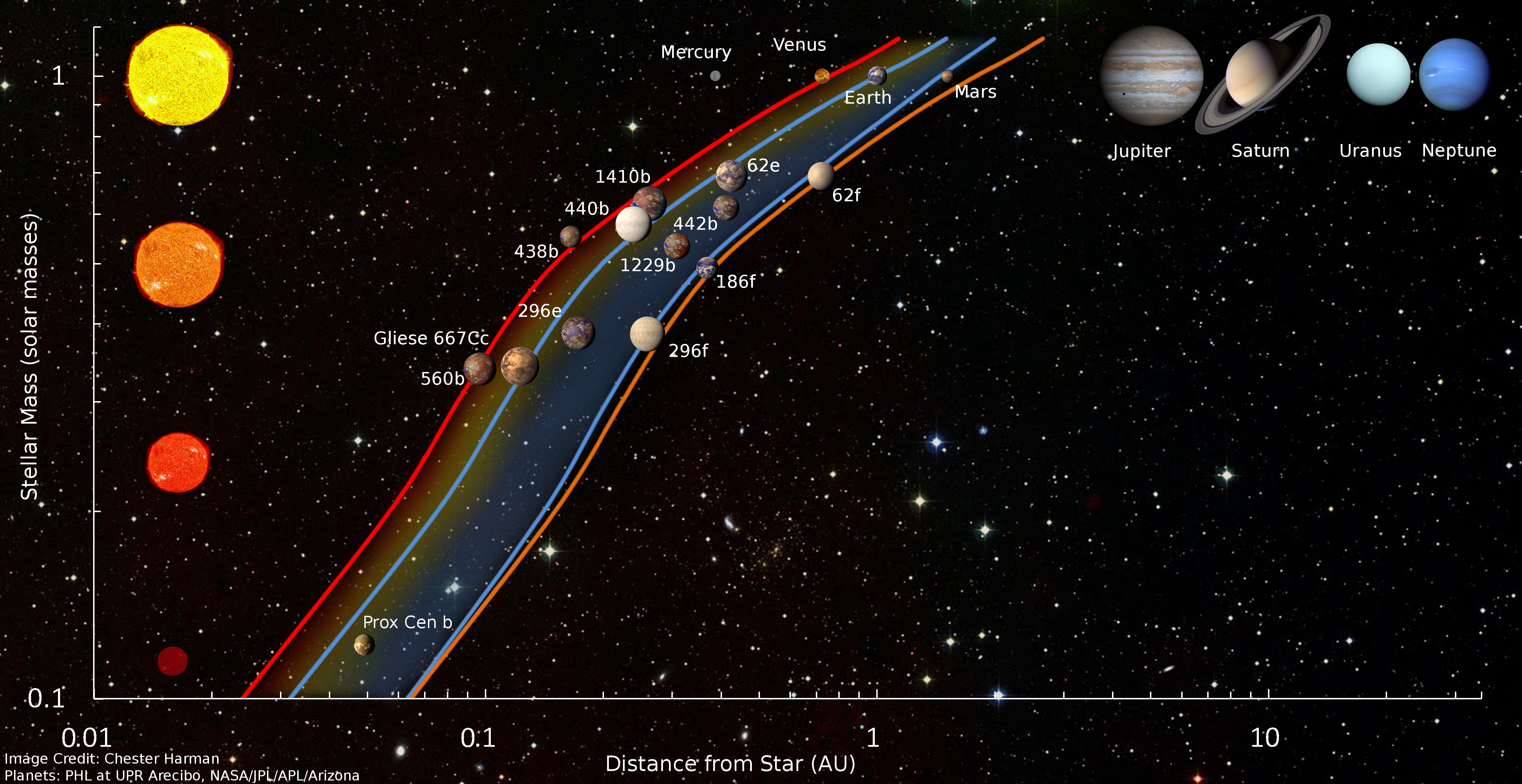}\\
\caption{Schematic diagram of the traditional orbital distance boundaries of the habitable zone (HZ) for stars of different masses. The red and orange lines denote the optimistic ``recent Venus" (inner) and ``early Mars" (outer) edges of the HZ. The blue lines bound the conservative ``runaway greenhouse" (inner) and ``maximum CO$_2$ greenhouse" (outer) HZ limits. The Solar System planets, as well as several habitable-zone exoplanets are plotted here for reference. This figure is reproduced with permission from the author. Image credit: Chester Harman.} 
\label{Figure 3. }
\end{center}
\end{figure}
    
Too close in to the star, and a planet risks losing its entire water inventory to space in a runaway greenhouse effect \citep{Ingersoll1969}, defining the conservative inner edge of the habitable zone (IHZ). At the outer edge of the habitable zone (OHZ) the maximum CO$_2$ greenhouse limit is reached. As regions of the spectrum become opaque, it becomes less effective to increase temperatures through the addition of CO$_2$ molecules into the atmosphere \citep{IPCC1990}. The effects of Rayleigh scattering then dominate over the greenhouse effect of increased CO$_2$, and temperatures above the freezing point of water can no longer be maintained by increasing CO$_2$ \citep{Kasting1993, Underwood2003, Pierrehumbert2010, Kane2012, Kopparapu2013b, Kopparapu2013a}. 

However, many factors and processes can affect planetary habitability other than orbital distance from a parent star. The long-term presence of surface liquid water depends upon the maintenance of a climate capable of sustaining the specific range of temperatures and pressures necessary to keep water in its liquid form. Over the past decade, the exploration and analysis of the potential habitability of extrasolar planets has expanded to include the use of multidimensional climate modeling. This approach has facilitated consideration of the broad range of stellar and planetary conditions that can influence the presence of surface liquid water on a potentially habitable world.
    
Planetary climate can be affected by the properties of the host star\textemdash its lifetime, activity level, and luminosity, among other factors\textemdash as well as the planet's individual properties, including its environment. An understanding of the interplay between stellar and planetary environments and their effects on planetary climate is crucial to an accurate assessment of the potential habitability of M-dwarf planets. 
    
 \subsection{Planetary Environment}   
A planet's climate is primarily determined by the incoming stellar radiation it receives from its host star, and the response of the planet's atmosphere and surface to that incoming energy, based on the planet's physical and orbital properties. Any imbalance due to the planet's response results in a change in surface temperature. Planetary surface temperature is therefore inextricably tied to the manner in which global energy balance is achieved between the incoming stellar energy and the outgoing thermal radiation emitted by the planet, which depends on the individual properties of the planet and its host star. 
    
The global energy balance of a planet is often expressed in terms of its ``equilibrium temperature"\textemdash the temperature of the planet assuming it is radiating back to space all of the incoming radiation it receives, as a blackbody. The presence of an atmosphere largely impacts the actual surface temperature on a planet, which, depending on the strength of its greenhouse effect, can be hundreds of degrees hotter than the equilibrium temperature, as is the case on Venus \citep{Seiff1987, Bougher1997}. Even with a relatively mild greenhouse effect, the presence and composition of an atmosphere can increase a planet's surface temperature by tens of degrees compared to its equilibrium temperature, as it does for the Earth \citep{Sagan1972}, which could mean the difference between a frozen planet and one that maintains an ample supply of liquid water on its surface. The equilibrium temperature therefore gives almost no information about the temperature on the surface of a planet with a significant atmosphere. While the identification of a planet orbiting its star at a sufficient distance to yield an equilibrium temperature between the freezing and boiling points of liquid water (273 K and 373 K, respectively) is a helpful first step, it by no means signifies the unequivocal discovery of a habitable planet. 
  
The specific composition of an atmosphere, and its interaction with the distribution of radiation from its parent star, can have a fundamental effect on planetary climate. In particular, atmospheric molecules that absorb strongly in the infrared, such as carbon dioxide (CO$_2$), water vapor (H$_2$O), methane (CH$_4$), and ozone (O$_3$) can absorb a larger fraction of the incoming stellar radiation (hereafter ``instellation") from a cooler star that has more of its output in the infrared (IR).

For planets orbiting M-dwarf stars, gases that absorb strongly in the IR and near-IR\textemdash where M dwarfs emit strongly\textemdash are most relevant. Previous work has shown that the broadband planetary Albedo\textemdash the reflectivity of the planet (including an atmosphere) integrated over the entire wavelength spectrum\textemdash decreases for simulated planets with Earth-like atmospheres around M-dwarf stars, compared to similar planets orbiting hotter, brighter stars. This is due in part to the larger absorption cross sections of CO$_2$ and water vapor in the near-IR, where M dwarfs are strongly emitting \citep{Kasting1993, Selsis2007, Shields2013, vonParis2013}. Additionally, the lower Albedo of surface ice and snow in the near-IR plays a role \citep{Joshi2012, Shields2013, vonParis2013, Shields2014}, as discussed in Section 5. This lower planetary Albedo may allow M-dwarf planets to absorb more instellation from their host stars, increasing their global mean surface temperature \citep{Kasting1993, Selsis2007, Shields2013, Shields2014, vonParis2013}. 

While habitability is often assumed to depend on the close alignment between a planet's atmospheric composition and that of the Earth, a wide range of atmospheric compositions and surface pressures is possible on exoplanets, and these different scenarios could have important implications for surface habitability on these worlds. For example, simulations of planets at a range of surface pressures have found that increased surface pressure results in larger horizontal heat fluxes, which reduce equator-to-pole temperature gradients, but smaller vertical heat fluxes, which cause increased surface temperatures \citep{Kaspi2015}. As will be discussed in Section 5, higher surface pressures could therefore increase the habitable surface area on synchronously-rotating planets or planets orbiting far from their stars \citep{Haberle1996, Joshi1997, Wordsworth2015}, by minimizing the amount of the planet that would otherwise freeze over entirely. 

The atmospheric constituents themselves will greatly influence resulting surface temperatures on potentially habitable planets. For planets with dense CO$_2$ atmospheres, atmospheric absorption of near-IR radiation will increase surface temperatures, which could increase the habitable surface area of a planet (e.g., \citealp{Wordsworth2010a, Wordsworth2011}), though CO$_2$ condensation will become more likely and collisional line broadening more critical above 1-2 bar of CO$_2$ \citep{Pierrehumbert2005}. Near the outer edge of their host stars' habitable zones, the effects of Rayleigh scattering will eventually dominate  over warming by atmospheric absorption of CO$_2$ \citep{Kasting1993, Selsis2007, Shields2013, Shields2014, Shields2016a}, resulting in frozen surface conditions. 

Alternative means of warming distant orbiting planets other than by increasing atmospheric CO$_2$ concentration have been explored, and would not require an active carbon cycle. Planets with thick envelopes of H$_2$\textemdash an incondensable greenhouse gas\textemdash may experience clement conditions for surface liquid water at far lower values of instellation than their counterparts with thick CO$_2$ atmospheres \citep{Pierrehumbert2011c}. However, thick hydrogen envelopes on planets at close distances to their stars could negatively impact habitability, by increasing surface temperatures to levels detrimental to surface life \citep{Owen2016}, though the high stellar activity of M dwarfs could possibly photoevaporate these envelopes, leaving habitable cores \citep{Luger2015a}. As will be discussed in Section 5, the intense stellar activity that is a unique property of M-dwarf stars, especially early in their lifetimes, can significantly affect atmospheric chemistry and evolution, particularly ozone column density \citep{Segura2005, Segura2010}, surface water concentration and atmospheric molecular oxygen abundance \citep{Luger2015b}, and atmospheric CO$_2$ inventory and stability \citep{Gao2015}, as well as the formation of photochemical cloud and haze layers \citep{Arney2016b, Meadows2016a}. These effects can impact the surface shielding of life from harmful UV radiation, the lifetimes and mixing ratios of biogenic gases such as methane (CH$_4$) in the atmospheres of M-dwarf planets \citep{Segura2005}, the spectroscopic detectability of biosignatures \citep{Meadows2016a}, and the likelihood of false positives for life \citep{Luger2015b, Gao2015}.  

Changes in ocean/land fraction have also been shown to affect planetary climate \citep{Dressing2010, Abe2011, Pierrehumbert2011b}. Global Climate Model (GCM) simulations of land planets orbiting Sun-like stars found these planets to be less susceptible to episodes of global-scale glaciation\textemdash so-called ``snowball" states, akin to the ``Snowball Earth" events \citep{Kirschvink1992} that may have occurred $\sim$720 and 635 million years ago on the Earth \citep{Pierrehumbert2011b}\textemdash than their ``aqua planet" (ocean-covered) counterparts, due to the lower thermal inertia of land and drier atmospheres \citep{Abe2011}. GCM simulations of M-dwarf planets have found that they may be less susceptible to snowball states regardless of land percentage \citep{Shields2013}. Additionally, while the traditional boundaries of the habitable zone are predicated on the assumption of a carbonate-silicate-cycle similar to that on Earth\textemdash where the silicate weathering rate increases with planetary surface temperature \citep{Walker1981}\textemdash the efficiency of a carbonate-silicate cycle on exoplanets remains unknown. Abbot \emph{et al.} (\citeyear{Abbot2012}) found that the weathering rate does not depend on land fraction on partially ocean-covered planets, as long as their surfaces are composed of at least 1\% land. However, using a plate tectonic-carbon cycle model, Foley (\citeyear{Foley2015}) found that the amount of exposed land area and total atmospheric CO$_2$ inventory influences the habitability of the steady-state climate on a planet. Hot, high-CO$_2$, ocean-dominated planets may lose all of their water before silicate weathering can re-stabilize the climate, and therefore may be less likely to exhibit clement conditions for life \citep{Foley2015}.


A planet's orbital elements and dynamics are essential to the discussion of climate and habitability. It has been long understood that at larger planetary obliquity\textemdash the angle between the planet's spin axis and the axis perpendicular to the orbital plane\textemdash a planet's seasonality increases \citep{Ward1974, Williams1975, Williams2003, Dobrovolskis2013}. Recent research has shown that habitable surface area increases for large obliquities \citep{Spiegel2009}. And planets that experience high-frequency oscillations in obliquity may avoid global glaciation, as neither pole of the planet faces away from the star for a long enough time for thick ice sheets to develop \citep{Armstrong2014}. A reduced negative cloud feedback on tidally-locked M-dwarf planets with non-zero obliquities could increase surface temperatures \citep{Wang2016}. The presence of moons may help stabilize a planet's obliquity \citep{Sasaki2014}, as is the case on the Earth \citep{Laskar1993}. However, the importance of a moon as a stabilizing mechanism remains uncertain. Recent work found that without the presence of a moon, any obliquity variations are significantly constrained \citep{Lissauer2012a}, and evolve slowly enough to eschew detrimental effects on long-term habitability \citep{Li2014}. Large moons could in fact adversely affect habitability on planets at the outer edge of a star's habitable zone, by preventing obliquity oscillations that may otherwise inhibit ice growth on a planet's surface \citep{Armstrong2014}. Additionally, annually-averaged insolation increases with orbital eccentricity \citep{Berger1993, Williams2002, Berger2006}, and modeling studies have shown that high-eccentricity planets exit episodes of global ice cover more easily \citep{Spiegel2010}. However, high-eccentricity, high-luminosity planets could experience dramatic climate differences between apoastron and periastron \citep{Bolmont2016b}, and planets with large eccentricities found in the habitable zone around low-mass stars could be subject to temperatures hot enough to negatively impact habitability \citep{Barnes2008}.

Statistical surveys indicate that $\sim$40\% of discovered planetary candidates are part of multiple-planet systems \citep{Rowe2014}, and that their false-positive discovery probability is low, implying that these candidates are likely to be planets \citep{Lissauer2012a, Lissauer2014}. The changes in spin state and orbital eccentricity possible as a result of interactions between a potentially habitable planet, its host star, and other planets in the system could have important consequences for climate and planetary habitability (e.g., \citealp{Shields2016a}). For example, planets with companions will have orbital eccentricities that undergo oscillations \citep{Mardling2007}, while also experiencing a strong tidal interaction from the parent star that can fix the rotation rate as a function of eccentricity. The particular stellar environment in which a planet resides can have a significant impact on habitability.
 
\subsection{Stellar Environment}
Although planetary habitability is affected by many other factors besides orbital distance from a parent star, the proximity of planets orbiting in the habitable zones of certain types of stars is an important consideration. Specifically, as discussed in detail in the following section, the habitability of planets orbiting M-dwarf stars is complicated by the close proximity of the habitable zones of these stars, given their lower stellar luminosities. In such close-in orbits, tidal forces between the star and orbiting planet are significant, and can modify the rotation rate of the planet.  

Planetary rotation rate has been shown to affect atmospheric circulation \citep{Joshi1997, Merlis2010, Showman2011a, Showman2011b, Showman2013, Kaspi2015}, as more slowly rotating planets exhibit a weaker Coriolis force, and longer periods of daytime illumination, which cause stronger convection and clouds in the substellar region of the planet \citep{Yang2014}, as shown in Figure 4. 

\linespread{1.15}
\begin{figure}
\begin{center}
\includegraphics [scale=0.5]{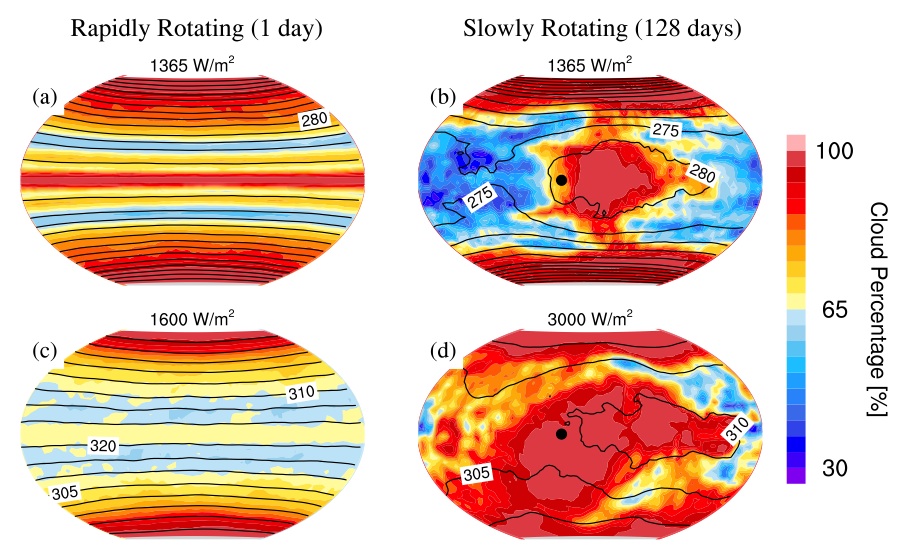}\\
\caption{Geographical differences in cloud percentage between rapidly- (left, 1 day) and slowly- (right, 128 days) rotating planets receiving different amounts of instellation, from Figure 2 of Yang \textit{et al.} (\citeyear{Yang2014}). Surface temperature in Kelvin is shown as black contours in intervals of 5 K. This figure is reproduced from \citet{Yang2014} with permission from the authors and AAS.} 
\label{Figure 4. }
\end{center}
\end{figure}

The tidal locking that can result from gravitational interactions between a star and a close-in planet can lead to captures into spin-orbit resonances \citep{Dole1964}, as has occurred for the planet Mercury \citep{Goldreich1966, Correia2004}. Synchronous rotation\textemdash a 1:1 spin-orbit resonance, and an extreme case of this tidal locking behavior\textemdash may occur, where the length of the planet's day is equal to its year. In this rotation state, one side of the planet faces the star in perpetual day, while the other side remains in eternal darkness. A planet whose substellar point is fixed is subject to enhanced radiative cooling \citep{Heng2012}, weakened low-latitude zonal winds (which also cool a planet), and increased atmospheric latent and oceanic diffusive heat transport, which could reduce day-night temperature differences on the planet \citep{Edson2011}.

Due to the relatively long eccentricity decay timescales \citep{Goldreich1966, Rasio1996} compared to the timescales for spin-orbit resonance capture \citep{Guillot1996, Rasio1996}, planets in a range of spin-orbit resonances with non-zero eccentricities could be common \citep{Wang2014a, Wang2014b}. Eccentric planets could exhibit vastly different global mean surface temperatures depending on the particular spin-orbit resonance and resulting pattern of insolation, which can affect planetary Albedo \citep{Wang2014a, Wang2014b}. Additionally, a planet with an oscillating eccentricity due to its companions can bounce chaotically between spin-orbit resonances \citep{Wisdom1984}, and the resulting effect on climate and habitability is unknown. 

The evolution in understanding of the extent to which tidal interactions between a star and a close orbiting planet can affect a planet's global energy budget has given rise to an expanded definition of the habitable zone beyond its traditional derivation. Recent studies have found that terrestrial exoplanets in eccentric orbits that take them into the habitable zones of low-mass stars could produce enough tidally-induced heating in their interiors to cause a runaway greenhouse state. This ``Tidal Venus" scenario could lead to planetary dessication, and is most likely to occur on planets orbiting stars of masses less than 0.3 M$_\odot$ \citep{Barnes2013}, underscoring the importance of including such mechanisms in the habitability appraisal of M-dwarf planets. However, tidal heating may also enhance habitability, by driving plate tectonics on small planets whose radiogenic heating is insufficient to the task, and could aid in maintaining planetary atmospheres through volatile outgassing \citep{Barnes2008, Jackson2008}.

The last decade of theoretical research has shed light on the limiting nature of the traditional habitable zone as it was originally devised. It is now understood that a detailed characterization of the planet host stars and their evolutionary histories is also essential to the discussion of which habitable-zone planets might sustain conditions that are conducive to the long-term survival of life \citep{Truitt2015}. It has become clear that both radiative and gravitational influences contribute strongly to the climatic evolution and stability of a planet, and by extension its ability to maintain liquid water on its surface over long timescales.  In the next section, we discuss in detail the radiative and gravitational effects that are unique to the M-dwarf stellar and planetary environment, and therefore crucial to the question of whether planets orbiting in the habitable zones of M-dwarf stars are likely to be habitable.

\section{New Theoretical Considerations and Their Impact on M-dwarf Planet Habitability}\label{sec5}
It was once believed that life could not survive on a planet orbiting an M-dwarf star. The environment would be too different from that of the Earth around the Sun given the harsh stellar radiation, short orbital periods and resulting strong tidal effects. Indeed, planets orbiting M-dwarf stars would be exposed to a very different environment compared to planets orbiting hotter, brighter stars. This is due to the unique radiative and gravitational properties of these small, cool stars. However, as understanding has grown regarding the differences between M-dwarf stars as a spectral type, and brighter, less long-lived stars, so too has consideration of the prospects for life on planets orbiting these cool, small stars. 

Questions have arisen about a planet's climate in the face of tidal effects induced between the planet and its close host star. The often tumultuous early lives of these stars have led to doubts about whether the very atmospheres of M-dwarf planets could withstand this prolonged phase of stellar evolution. And M-dwarf protostellar nebulae\textemdash the natal origins of these small red stars\textemdash have proven relevant to the search for liquid water on the surfaces of these worlds, and vital to the discussion of whether M-dwarf planets as a class are expected to be water rich. 

It is through a combination of employing observational data for recently discovered planetary systems, and theoretical simulations using a hierarchy of computer models, that a deepened understanding has emerged of the effect on planetary climate and habitability of the wide range of properties that govern a planet's ability to sustain liquid water on its surface (see Section 4). This unified approach has allowed the identification of those planets that demonstrate the greatest likelihood for habitability across the broad range of factors that are currently unconstrained, and for which observational data are still limited. This approach has also improved our understanding of the likelihood of planets orbiting certain spectral classes of stars to be habitable.

Modeling efforts have concentrated on exploring the variations in these radiative and gravitational properties to quantify their effect on planetary climate, particularly with regard to planets orbiting M-dwarf stars. In this section we summarize the chief contributions to the field of knowledge regarding theoretical considerations crucial to the habitability of these planets, focusing on computationally-driven advances in research over the past decade.   

\subsection{Radiative Effects}
The radiative environment surrounding an M-dwarf star is very different from that of a hotter, Sun-like or brighter star. This is due to the lengthy lifetimes of M-dwarf stars compared to more luminous stars with rapid fuel-burning rates. The slower fuel-burning rate for M-dwarf stars is a chief advantage for habitability, because such stars offer abundant timescales for planetary and biological evolution. The earliest fossil evidence on Earth suggests that it took $\sim$0.8-1 Gyr for life to develop on the Earth \citep{Awramik1992, Mojzsis1996, McKeegan2007, Nutman2016}, though recent results suggest that signs of the emergence of life date back as far as 0.5 Gyr after the Earth's formation \citep{Bell2015}.

However, the long lifetimes of M-dwarf stars mean that these stars can be extremely active, particularly early in their lifetimes. The fraction of stars that are magnetically active has been found to increase monotonically with later spectral type\textemdash likely due to the longer activity lifetimes of late-type M dwarfs \citep{West2008}\textemdash and reaches a peak at around a spectral type of M7 \citep{Hawley1996, Gizis2000, West2004}. A strong correlation exists between stellar rotation period and magnetic activity in early-type M dwarfs \citep{Mohanty2003}, while it is less pronounced (though still present) in fully convective, late-type M dwarfs \citep{West2015}. The intense chromospheric activity exhibited by M dwarfs can result in flares \citep{Hawley1991, Lammer2013}, which have been found to be more frequent \citep{Hilton2010b, Davenport2012} and with larger amplitude \citep{Hilton2011} for stars of later spectral type. Even less active M dwarfs have been found to exhibit significant X-ray and UV (XUV) emission and flare activity \citep{France2013, France2016}. 

The extreme XUV radiation from flares is particularly relevant for planetary habitability. M-dwarf stars have an extended pre-main sequence (PMS) phase \citep{Baraffe1998, Baraffe2015}\textemdash the period of time prior to settling onto the Main Sequence to begin hydrogen fusion in their cores. This could present significant challenges for orbiting planets. While planets orbiting older M-dwarf stars may receive similar XUV emission levels to that received by the Earth \citep{Guinan2016}, M-dwarf stars exhibit saturated emission levels for the first 0.5\textendash1 Gyr, or longer for later-type M stars \citep{Lammer2009}. The stellar activity of M dwarfs is high, resulting in large amounts of XUV radiation emitted toward the surface of the planet, which may cause atmospheric erosion \citep{Lammer2007b}, a runaway greenhouse, and hydrodynamic escape on close orbiting planets \citep{Luger2015b}. Over the $\sim$1 Gyr of an M-dwarf star's intense activity, a planet orbiting in its habitable zone could have been bombarded with this XUV radiation. By the time the M dwarf has settled onto the Main Sequence, planets that were once habitable may have lost oceans worth of water to space, and could be long dessicated and void of surface life \citep{Luger2015b}. However, a recent study of possible water loss in the atmospheres of ultracool dwarfs (T$_{eff}$ $<$ $\sim$ 3000 K) found that the TRAPPIST-1 planets \citep{Gillon16}, particularly TRAPPIST-1d, may still have retained enough water to maintain surface habitability, depending on their original water inventories \citep{Bolmont2016a}. 

M-dwarf planets observed today orbiting well outside of the inner edge of their stars' habitable zones may have been significantly interior to the runaway greenhouse distance threshold for 1 Gyr or more, as shown in Figure 5. Over time the atmospheres of these planets could become oxygen rich (Figure 5), due to the evaporation of oceans into the stratosphere, subsequent water photolysis, and the escape of the lighter H$_2$ to space (\citep{Luger2015b}, also see \citep{Ramirez2014} and \citep{Tian2015}). A low inventory of non-condensing gases such as N$_2$ in a planet's atmosphere could prevent cold trapping of H$_2$O in the upper atmosphere, increasing the likelihood of abiotic O$_2$-rich worlds \citep{Wordsworth2014}. Depending on the atmospheric hydrogen content, a subsequent CO$_2$-dominated atmosphere may be depleted in great enough quantities to generate Earth-like abundances of abiotically produced O$_2$ and O$_3$ \citep{Gao2015}. This could provide a false-positive signal observationally for life \citep{Gao2015, Luger2015b}, though it may be possible to distinguish these planets from planets with biogenic O$_2$ in their atmospheres \citep{Gao2015, Schwieterman2016a, Schwieterman2016b}. Additionally, due to the larger ratio of far-UV to near-UV radiation for M dwarf stars \citep{France2013, France2016}, CO$_2$-rich M-dwarf planets may have a greater susceptibility to develop abiotic O$_2$-rich atmospheres created by CO$_2$ photolysis (followed by recombination of O atoms with other O atoms) \citep{Tian2014, Harman2015}, particularly if O$_2$ surface sinks are small. 

\linespread{1.15}
\begin{figure}
\begin{center}
\includegraphics [scale=0.4]{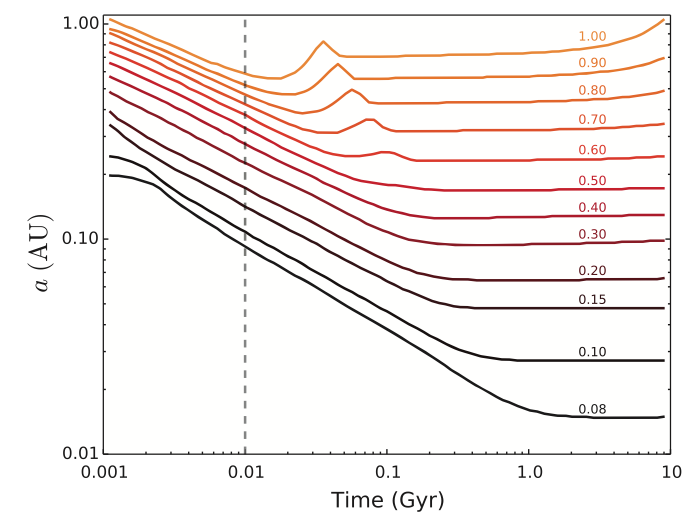}
\includegraphics [scale=0.3]{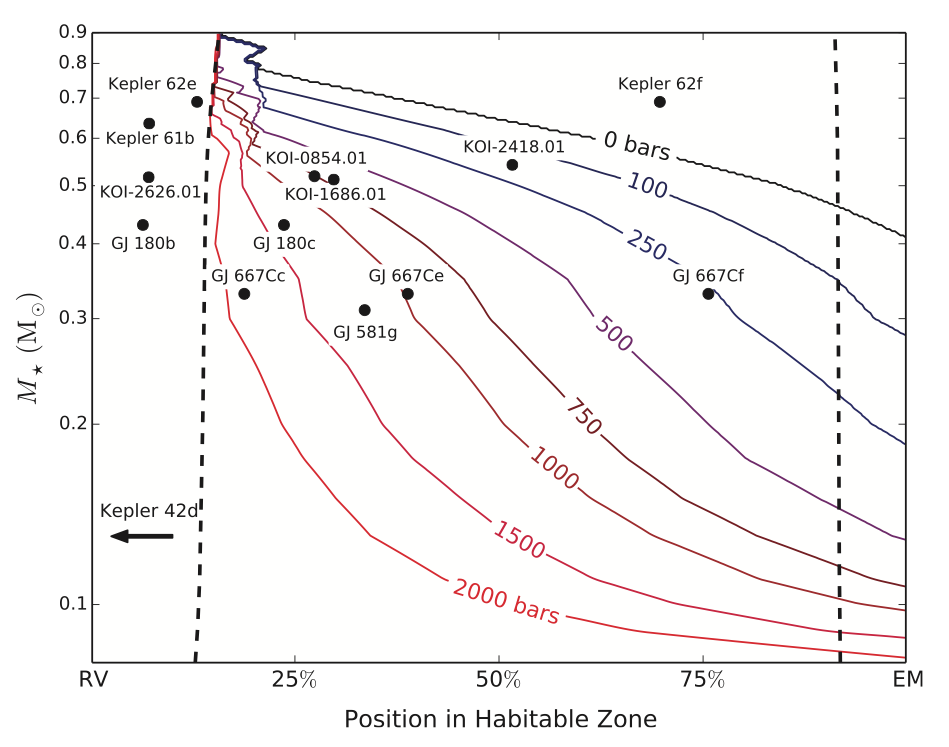}\\
\caption{\textbf{Top}: The location of the empirical inner edge of the habitable zone (Recent Venus limit) as a function of time for stars from 0.08M$_\odot$ to 1M$_\odot$, from Figure 3 in Luger and Barnes (\citeyear{Luger2015b}). The vertical dashed line denotes the 10 Myr formation timescale assumed in calculations. \textbf{Bottom}: From Figure 13 of in Luger and Barnes (\citeyear{Luger2015b}), a selection of recently-discovered super Earths that could have detectable O$_2$ atmospheres if they formed with a significant amount of surface water. The vertical dashed lines denote the Runaway Greenhouse (inner edge) and Maximum CO$_2$ greenhouse (outer edge) limits of the habitable zones on the left and right, respectively. Calculations assume a mass of 5M$_\oplus$, initially 10 terrestrial oceans' worth of surface water, and diffusion-limited escape. Contours show the atmospheric pressure of O$_2$ in bars that could have built up by the end of the runaway greenhouse phase, assuming all of it remained in the atmosphere. This figure is reproduced from \citet{Luger2015b} with permission from the authors and Mary Ann Liebert, Inc.} 
\label{Figure 5. }
\end{center}
\end{figure}

For planets that avoid the fate of losing their oceans, the high stellar activity of M-dwarf stars presents complications for their atmospheres, and for surface life. The effect of the UV radiation environment of M-dwarf host stars on the atmospheric photochemistry of orbiting planets has been recently explored \citep{Segura2010, Rugheimer2015a, Rugheimer2015b}. Simulations of the incident UV flux from mid-M dwarf AD Leo found that the ozone number density within the atmosphere of an orbiting M-dwarf planet may not be significantly reduced by more than 1\%, unless the proton flux that sometimes accompanies flare events is included, in which case the atmospheric ozone column depth could be depleted by as much as 94\% \citep{Segura2010}. Indeed, assuming flare characteristics similar to those of an active M dwarf such
as GJ1243 \citep{Hawley2014} and an Earth-like planet orbiting in the habitable zone, the flare activity and associated energetic proton events could strongly impact the ability of the planet's atmosphere to shield the surface from harmful UV radiation, even for infrequent ($\sim$1 per week) repeated proton events (M. Tilley, private communication). However, these ionizing particles are not always present during flare events. Additionally, recent theoretical work found that more than 25\% of the ozone column remains on an M-dwarf planet that is subjected to high stellar activity \citep{Tabataba-Vakili2016}. This is due to the cosmic ray-induced production of HO$_\textit{x}$ species (OH, HO$_2$), which react with NO$_\textit{x}$ species (NO and NO$_2$) to produce HNO$_3$, thereby offsetting the destruction of ozone by NO$_\textit{x}$. The result is a residual (albeit weaker) ozone signal in modeled spectra \citep{Tabataba-Vakili2016}, which could be a contributing biosignature if detected along with oxygen \citep{DesMarais2008}. And this combination would be more easily detected given early, active (rather than later-type) M-dwarf host stars, due to the increased UV fluxes \citep{Rugheimer2015b}. The primary CO$_2$-dominant atmospheres of super Earths in the habitable zones of M-dwarf stars may be more stable against thermal escape induced by XUV radiation, and retain more oxygen due to the preferential escape of CO$_2$ \citep{Tian2009}. 

The detection of biosignatures\textemdash biologically-generated global impacts to a planet's atmospheric and/or surface environment that could be observed remotely \citep{Meadows2005, Meadows2016a}\textemdash would offer important clues about the habitability and evolutionary path of life on an M-dwarf planet. To place biosignatures in environmental context, indications that the planet is habitable should also be sought. The identification of specular reflection or ``glint" from starlight on the surface of a planet, while challenging, could indicate the presence of an ocean \citep{Williams2008, Robinson2010, Robinson2014, Meadows2016a}, where life may have originated and developed. However, false positives for ocean glint are possible \citep{Cowan2012}, and any remotely observed specular reflection from a surface ocean may not necessarily indicate the presence of liquid water, as evidenced by similar observations of hydrocarbon lake regions on Saturn's moon Titan \citep{Stephan2010}. Changes in surface reflectivity as a planet rotates beneath an instrument's viewing angle could be used to distinguish between the presence of oceans, ice, continents, or vegetation \citep{Cowan2009, Cowan2011, Robinson2011}. The vegetation ``red edge" on Earth at wavelengths near 0.7 $\mu$m \citep{Gates1965, Seager2005, Kiang2007a} can potentially be detected in the Earth's disk-averaged spectrum \citep{Tinetti2006b, Montanes-Rodriguez2006}. If measured spectroscopically in a planet's atmosphere, O$_2$, or O$_3$, along with methane (CH$_4$)\textemdash which has been proposed to be more abundant and long lasting in the atmospheres of M-dwarf planets, due to the lower amounts of near-UV surface emission from these stars \citep{Segura2005}\textemdash could be indicative of a possibly biogenic source of O$_2$, such as oxygenic photosynthesis \citep{Lovelock1965}. However, much recent work has been done to identify abiotic sources of O$_2$ on M-dwarf planets \citep{Tian2014, Wordsworth2014, Domagal-Goldman2014, Luger2015b, Gao2015, Harman2015, Tian2015, Schaefer2016, Barnes2016}, and to also understand how to discriminate abiotic from biological sources using spectroscopic observations \citep{Domagal-Goldman2014, Gao2015, Harman2015, Schwieterman2015, Schwieterman2016a, Schwieterman2016b, Meadows2016a}. Methane, in addition to its biological sources, can be produced by abiotic processes such as serpentinization (the alteration of rocks with the addition of water into the crystal structure of their minerals; e.g. \citealp{Kelley2005, Etiope2013}), although this water- and tectonically- driven process itself could be a sign of habitability \citep{Arney2016a, Meadows2016a}. The detection of high concentrations of O$_3$ would increase the likelihood that any surface life would be protected from significant UV flare-induced biological damage \citep{Segura2010} by a thick ozone layer, potentially affording life the opportunity to develop on land. However, a thin or non-existent ozone shield may not preclude the development of life on an M-dwarf planet, since the ocean could also provide UV radiation shielding \citep{Kiang2007b}. Additional biosignatures such as nitrous oxide (N$_2$O, \citealp{DesMarais2002}), methyl chloride (CH$_3$Cl, \citep{Segura2005}), and ethane, produced from photolysis of methyl-bearing sulfur-rich gases like dimethyl sulfide (DMS) and dimthyl disulfide (DMDS) could possibly be measured in the atmospheres of M-dwarf planets with inactive (and therefore low UV flux-emitting) host stars, and could indicate the presence of microbial life \citep{Domagal-Goldman2011}. 

The ultimate effect of flare activity from an M dwarf could be weakened due to the protection of a planet's magnetic field \citep{Segura2010}, though its strength on M-dwarf planets is uncertain. The strength of a planet's magnetic field is critical to its ability to maintain an atmosphere\textemdash that crucial shield that helps regulate a planet's climate, surface temperature, and day-night temperature contrasts \citep{Ward2000}. Planetary magnetic fields also help protect the surface from cosmic-ray particles and other forms of incoming stellar radiation that may be harmful to biology \citep{Griessmeier2005, Dartnell2011, Griessmeier2015}. Planetary magnetic moments may be weakened on tidally-locked M-dwarf planets \citep{Lammer2007a, Khodachenko2007}. Recent work found that magnetic field strengths of tens to hundreds of Gauss may be required to protect the atmospheres of habitable-zone planets orbiting mid M-dwarf stars from coronal mass ejections, or CMEs \citep{Kay2016}. While it could be challenging for rocky exoplanets to attain that strong of a magnetic field, it would likely be easier for similar planets orbiting early M dwarfs to retain their atmospheres, given the lower expected CME impact rates \citep{Kay2016}. Studies have found that tidally-induced heating of an Earth-mass planet with a fixed, eccentric orbit in the habitable zone of a low-mass star can cause a weakened magnetic field, among other effects, if the tidal heating is strong enough \citep{Driscoll2015}. Driscoll and Barnes (\citeyear{Driscoll2015}) also found that magnetic fields around M-dwarf planets should be common.  

A planet's magnetic field also offers crucial protection from stellar wind erosion of its atmosphere \citep{Driscoll2013}. The stellar wind dynamic pressures of M-dwarf stars could be several orders of magnitude greater than those of the Sun, and nonuniform (see e.g., \citealp{Garraffo2016}), resulting in a significant amount of Joule heating at the top of the atmosphere of close-in, habitable-zone planets \citep{Cohen2014}. Stellar winds and extreme UV (EUV) activity have been shown to have an insignificant effect on the atmospheric mass loss rate of Venus-like M-dwarf planets, unless atmospheric ion acceleration is present \citep{Cohen2015}. However, stellar rotation rates of mid-to-late M-dwarf planets may be so fast that the magnetic fields of these stars would reduce the size of the magnetospheres of Earth-sized planets, resulting in stellar wind erosion of planetary atmospheres \citep{Vidotto2013}.

If the magnetic fields of M-dwarf planets are weakened compared with those of similar planets orbiting Sun-like or brighter stars, recent work has found that surface biology may not be affected. While stratospheric ozone concentration may be reduced by as much as 20\% by galactic cosmic-ray protons, recent work has found that the increased UV flux reaching the planet's surface would have a negligible effect on surface biology, assuming an atmosphere with a surface pressure equal to that of the Earth \citep{Griessmeier2016}.

Recent statistical studies of Kepler planets have found that a significant fraction of close-in, small planets have large H/He gas envelopes, which could preclude habitability, by increasing temperatures and pressures outside of the range for surface liquid water \citep{Owen2016}. Theoretical studies have found that photoevaporation of these gas envelopes would be possible, given the intense XUV activity inherent to M dwarf stars \citep{Hawley1991, Scalo2007, Segura2010, Lammer2013}, and that these planets could lose as much as $\sim$1\% of the H/He envelope, yielding habitable planets if the planets' cores are small enough ($M\leqslant$0.9M$_\oplus$ near the IHZ or $M<$0.8M$_\oplus$ near the OHZ), and hydrodynamic escape is the chief mechanism \citep{Owen2016}.

It is difficult to directly measuring the Lyman-alpha (Ly$\alpha$) emission line at 1216\AA\textemdash which dominates the Far-UV flux for late (cooler) M dwarfs\textemdash due to heavy attenuation by neutral hydrogen in the interstellar medium \citep{France2013, Linsky2013}. However, recent studies exploring the correlation between Ly$\alpha$ emission, and emission at other wavelengths more easily detected, using measurements from the \textit{Hubble Space Telescope} \citep{Linsky2013} and the \textit{Galaxy Evolution Explorer} \citep{Shkolnik2014b}, will increase the accuracy of photochemical models of exoplanet atmospheres, and help constrain the effect of the UV environment of M dwarfs on the atmospheric evolution of orbiting habitable-zone planets.

One of the areas of research that has seen the most marked evolution over the last decade is the exploration of the interaction between the spectral energy distribution (SED) of M-dwarf stars and the atmospheres and surfaces of their orbiting planets. The warming caused by atmospheric CO$_2$ and water vapor absorbing strongly in the near-IR, where M dwarfs emit strongly, has been established for some time, based on earlier work \citep{Kasting1993, Selsis2007}. And recent studies have continued to explore the effect of stellar SED on the vertical temperature structure of an orbiting planet's atmosphere, and its resulting climate \citep{Shields2013, vonParis2013, Godolt2015}. What has only recently emerged is an understanding of the influence on climate of the interaction between the SED of an M-dwarf host star and the icy and snowy surfaces of an orbiting planet. 

When considering the effect of ice-Albedo feedback on planetary climate, most often what is considered is the interaction between ice and our own host star, the Sun. Ice and snow largely appear bright on the Earth, and the more light ice reflects back to space, the less it absorbs. This leads to cooler temperatures and more ice growth, resulting in increased solar radiation reflected away from the planet via this positive feedback process.

However, ice and snow have a spectral dependence, therefore the ice-Albedo feedback mechanism is sensitive to the wavelength of light coming from the host star. This means that the interaction between an M-dwarf SED\textemdash which peaks in the near-IR\textemdash and a planet's atmosphere and surface will have a different climatic effect from that between the SEDs of other host stars with more visible and near-ultraviolet (near-UV) output and their orbiting planets. 

Figure 6 shows the SEDs of different spectral classes of stars, from an F-dwarf star (hotter and more luminous than the Sun) to an M-dwarf star.  Over 95\% of the M-dwarf star's radiation is emitted at wavelengths longer than 1 $\mu$m, compared to 53\% for the Sun. While the Albedo (reflectance) of ice and snow is high in the visible and near-UV, where G- and F-dwarf stars emit strongly, in the near-IR the reflectance of ice and snow drops substantially. This drop in reflectance is due to molecular vibrations from combinations of the water molecule's three vibrational modes \citep{Farrell1967}, resulting in an increase in the absorption coefficients of ice and snow at larger wavelengths \citep{Dunkle1956}. This means that on M-dwarf planets, a large fraction of the instellation that reaches the surface of the planet will be absorbed by, rather than reflected from surface ice and snow. As a result, the ice-Albedo feedback mechanism, which is largely cooling on the Earth, may be suppressed on M-dwarf planets \citep{Joshi2012}. 

\linespread{1.15}
\begin{figure}
\begin{center}
\includegraphics [scale=2.00]{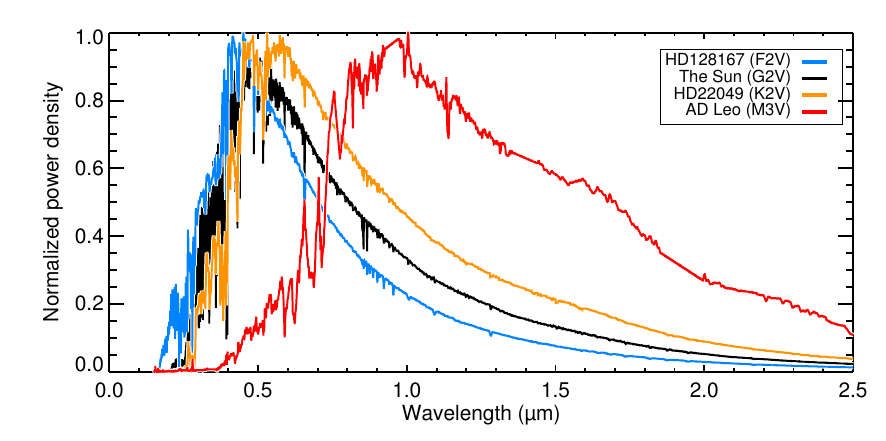}\\
\caption{The SEDs for F-, G-, K-, and M-dwarf stars, normalized by their peak flux, as shown in Figure 1 of Shields \textit{et al.} (\citeyear{Shields2013}). This figure is reproduced from \citet{Shields2013} with permission from the authors and Mary Ann Liebert, Inc.} 
\label{Figure 6. }
\end{center}
\end{figure}

Comprehensive follow-up studies of this phenomenon carried out with a hierarchy of radiative transfer and climate models found that lower-mass planets exhibit warmer climates than planets orbiting hotter, more luminous stars at equivalent stellar flux distances \citep{Shields2013, vonParis2013}. M-dwarf planets appear less susceptible to snowball states, requiring a larger decrease in instellation for global glaciation \citep{Shields2013}. This lower climate sensitivity to changes in instellation for M-dwarf planets is due to both ice and snow absorbing strongly in the near-IR, and atmospheric absorption, which reduces the temperature lapse rate throughout most of the atmospheric column, resulting in more shortwave (incoming stellar) radiation absorbed by the planet. Indeed, as recent work simulating land planets orbiting G-dwarf stars found that they froze over at 77\% of the modern solar constant, compared to 90\% instellation for G-dwarf aqua planets \citep{Abe2011}, M-dwarf aqua planets appear to require larger decreases in instellation to freeze regardless of land percentage \citep{Shields2013}. As Hadley circulation on deglaciating simulated M-dwarf planets has been shown to be weaker, M-dwarf planets may also thaw more easily out of global ice cover, as indicated by the smaller hysteresis for M-dwarf planets compared with planets orbiting G- and F-dwarf stars in Figure 7. M-dwarf planets may therefore have more stable climates over long timescales, which could make them more amenable for the development and evolution of life \citep{Shields2014}.

\clearpage
\linespread{1.15}
\begin{figure}
\begin{center}
\includegraphics [scale=0.6]{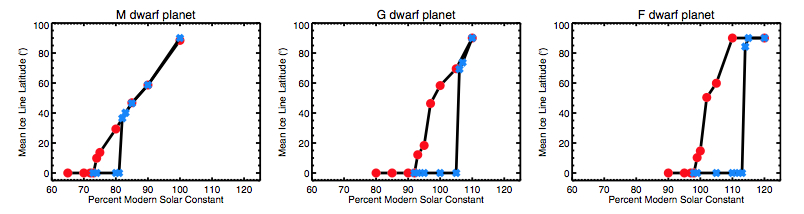}\\
\caption{Mean ice line latitude as a function of stellar flux for an aqua planet orbiting an M-, G-, and F-dwarf star, as shown in Figure 1 of Shields \textit{et al.} (\citeyear{Shields2014}). Simulations assuming an initial warm start are in red (circles). Initial cold start simulations are in blue (asterisks). This figure is reproduced from \citet{Shields2014} with permission from the authors and AAS.} 
\label{Figure 7. }
\end{center}
\end{figure}

Originally it was proposed that the traditional limit of the outer edge of the habitable zone for M-dwarf stars could be extended because of the lower Albedo of planetary surface ice and snow in the near-IR \citep{Joshi2012}. However, if an active carbonate-silicate cycle exists on these planets, where CO$_2$ increases in the atmosphere as the silicate weathering rate decreases due to lower temperatures \citep{Walker1981}, high CO$_2$ would be expected to build up in a planet's atmosphere at farther distances from its star. Additional work done with computer models found that a planet's sensitivity to the ice-Albedo effect depends on atmospheric composition and surface pressure, and that for dense CO$_2$ atmospheres, atmospheric properties rather than surface properties dominate the planetary energy balance and radiation budget \citep{Shields2013, vonParis2013}. Ultimately, the outer edge of the habitable zone for M dwarfs largely remains unaffected by the spectral dependence of snow and ice Albedo, as high atmospheric CO$_2$ would likely mask the planet's climate sensitivity to the surface, and to the ice-Albedo effect. However, as M-dwarf planets exhibit stronger radiative responses to increases in CO$_2$ in climate simulations, they may require less CO$_2$ to build up farther away from their stars to maintain warm enough surface temperatures for surface liquid water \citep{Shields2013}. The habitable zone could be reduced by limited CO$_2$ outgassing \citep{Kadoya2014, Abbot2016}, but this reduction would likely be less for M-dwarf planets, given their lower planetary Albedos \citep{Abbot2016}. Additionally, the amount of CO$_2$ generated on synchronously-rotating planets could be much higher if the substellar point is located in a region with an ocean versus a land-covered area, creating an even stronger radiative response \citep{Edson2012} on planets orbiting M-dwarf stars.

The likelihood of the process of photosynthesis to operate on M-dwarf planets has been discussed for some time, given the shift in the peak wavelength to the near-IR for M-dwarf stars, compared to the visible for Sun-like, G-dwarf stars \citep{Kiang2007a, Kiang2007b}. Since photosynthesis occurs at visible wavelengths on the Earth, the paucity of incident radiation in either wavelength range in an M-dwarf spectrum implied that plants would have fewer photons for oxygenic photosynthesis. However, as discussed in Section 6, recent work has found that both oxygenic, as well as anoxygenic photosynthesis (in a low-O$_2$ atmospheric scenario) could indeed occur on M-dwarf planets. Underwater niches, in addition to providing protection from biologically harmful EUV radiation, could permit access to sufficient light for organisms to use for chemical processes and growth \citep{Kiang2007b}.

\subsection{Gravitational Effects}

A chief concern regarding the habitability of M-dwarf planets arises from the close proximity of these planets orbiting in the habitable zones of M-dwarf stars, which are much closer in to the stars than those of G- or F-dwarf stars \citep{Kopparapu2013c}. In such close-in orbits, tidal effects are expected to be strong, and could result in capture into resonances in spin-orbit period \citep{Dole1964}, as has occurred for Mercury \citep{Goldreich1966, Correia2004}. Synchronous rotation\textemdash an extreme case of tidal locking\textemdash is possible, where the planet's orbital period is equal to its rotation period. Such a state would leave one side of the planet always facing the star in perpetual daytime, while the other side remains in eternal darkness. This has raised concerns about the possibility of the atmospheres condensing out onto the night sides of synchronously-rotating M-dwarf planets. 

However, these concerns have been largely ameliorated through significant modeling efforts in this area. The weaker Coriolis force present on synchronously-rotating planets will result in advection dominating as the primary means of transporting heat to the night side of the planet \citep{Showman2013}. This result along with previous work has shown that planets with at least 0.1 bar of a greenhouse gas such as CO$_2$ in their atmospheres could distribute enough heat to the night side of the planet to avoid atmospheric freeze-out \citep{Haberle1996, Joshi1997, Wordsworth2015}, resulting in a smaller surface temperature contrast between the day and night sides of the planet \citep{Pierrehumbert2011a, Wordsworth2011}. Ocean heat transport can increase the efficiency of this process \citep{Hu2014}. Synchronously-rotating planets receiving lower amounts of instellation may have less stringent atmospheric constraints for heat redistribution \citep{Goldblatt2016}.

Additionally, thermal atmospheric tides raised on a close orbiting planet could be significant compared to gravitational tides \citep{Gold1969, Ingersoll1978, Leconte2015, Cunha2015}. Recent work exploring the effect of thermal tides in the atmospheres of M-dwarf planets indicates that for high stellar insolation, an atmospheric torque may be generated that opposes tidally-induced spin synchronization, as is the case for Venus. This implies that M-dwarf planets may not necessarily be synchronously rotating in their host stars' habitable zones \citep{Leconte2015}. 

Additionally, a climatic benefit from synchronous rotation has been recently proposed for M-dwarf planets orbiting near the IHZ. Studies have shown that cloud cover fraction is a function of planetary rotation rate, with larger cloud cover created on planets (with oceans) rotating at slower periods \citep{Yang2014}, as shown in Figure 8 for tidally vs. non-tidally locked planets.  On simulated synchronously-rotating planets, increased cloud cover is formed at the substellar point (providing it is located over an ocean), reflecting incident stellar flux, and buffering these planets against runaway greenhouse states compared to planets whose substellar point varies greatly with time \citep{Yang2013}. This would extend the habitable zone for M-dwarf stars, as shown in Figure 9. However, these results have been shown to be sensitive to planetary orbital and rotational period, as well as stellar metallicity \citep{Kopparapu2016}. 

\linespread{1.15}
\begin{figure}
\begin{center}
\includegraphics [scale=0.6]{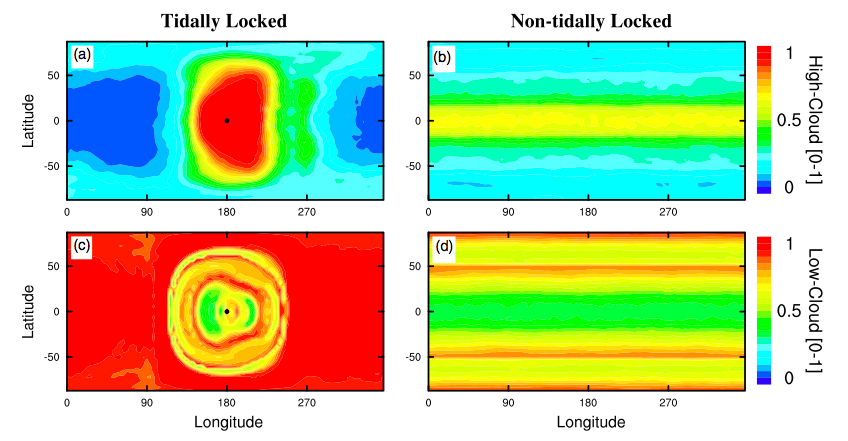}\\
\caption{Figure 1 from Yang \textit{et al.} (\citeyear{Yang2013}) showing cloud behavior for synchronously- (left) and non-synchronously rotating planets (right). High-level (top) and low-level (bottom) The non-cloud fraction are shown for each case. The non-synchronous case is in a 6:1 spin-orbit resonance. The instellation received by the planet is 1400 W/m$^2$. The black dots in (a) and (c) identify the substellar point on the synchronously rotating planets. This figure is reproduced from \citet{Yang2013} with permission from the authors and AAS.} 
\label{Figure 8. }
\end{center}
\end{figure}

\linespread{1.15}
\begin{figure}
\begin{center}
\includegraphics [scale=0.6]{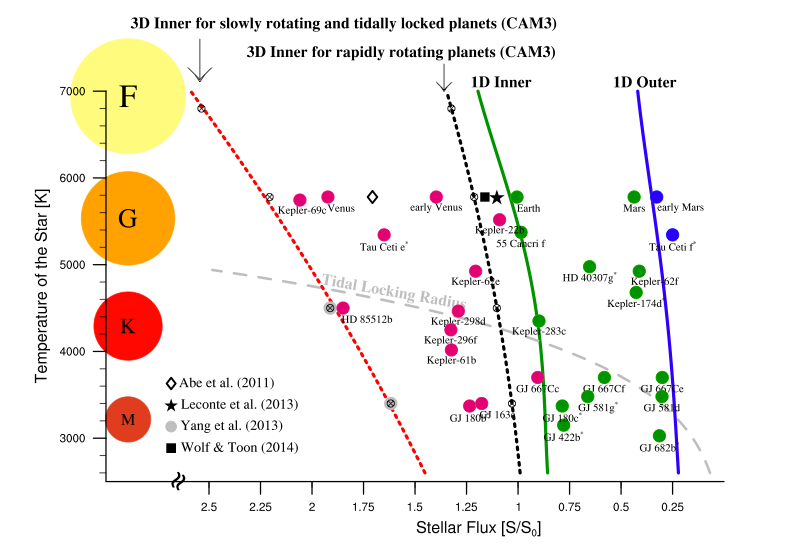}\\
\caption{Habitable zone boundaries as a function of stellar type and planetary rotation rate derived using 1-D radiative-convective model and the 3-D General Circulation Model CAM3, from Yang \textit{et al.} (\citeyear{Yang2014}). Green and blue lines denote the inner (runaway greenhouse) and outer (maximum CO$_2$ greenhouse) edges of the habitable zone, respectively, using the 1-D model \citep{Kopparapu2013a,Kopparapu2013b}. The black and red lines indicate the inner edge of the habitable zone calculated using CAM3 and assuming a rapidly- (1 day) and slowly- (128 days orbiting G and F stars, tidally locked with a 60-day orbital period for M and K stars) rotating planets. The grey line is the tidal locking radius \citep{Kasting1993}. Also plotted are the inner edge limit for a dry planet orbiting a G-dwarf star \citep{Abe2011}, and for the Earth obtained with LMD Generic GCM \citep{Leconte2013b}, and using a modified radiative transfer code \citep{Wolf2014}. Solar system planets and recently discovered exoplanets are plotted with for reference. This figure is reproduced from \citet{Yang2014} with permission from the authors and AAS.} 
\label{Figure 9. }
\end{center}
\end{figure}

Recent work explored the effects of tides on the internal heat budget of Earth-mass planets in the habitable zones of M-dwarf stars \citep{Barnes2008, Barnes2013, VanLaerhoven2014, Driscoll2015}. Given the long ages of M dwarfs, Earth-sized planets orbiting in their habitable zones may have cooled significantly over their planetary and stellar lifetimes, rendering plate tectonics\textemdash a geochemical recycling process crucial to climate and habitability on Earth for its enabling of the carbonate-silicate cycle \citep{Foley2015}\textemdash inert. A planet could remain warm enough for geophysical activity, and perhaps life, after its internal heat sources have ceased, due to tidally-induced heating \citep{Barnes2008}, which could be caused by the presence of an outer companion \citep{VanLaerhoven2014}. This could prolong habitability for old, small planets orbiting low-mass stars. Since tidal forces between a close-in planet and the star can result in non-negligible changes in orbital distance, M-dwarf planets found interior to the inner edge of the habitable zone today may have been comfortably within the limits of their stellar habitable zones in the past, and could have been habitable \citep{Barnes2008}. However, for planets in fixed eccentric orbits around stars with masses less than $\sim$0.45 M$_\odot$, strong tidal forces could also lead to a weakened planetary magnetic field, high volcanic activity, and prolonged magma oceans, all of which could cause a dessicated or otherwise inhospitable surface environment for life \citep{Driscoll2015}. 

When discussing surface habitability, the implications of an ocean are crucial, given the assumption of surface liquid water as the most important indicator of a habitable planet. The building blocks of a habitable planet\textemdash including an abundance of volatiles, such as water, in the planet-forming disk material from which a terrestrial planet emerges\textemdash are therefore essential. Research has shown that the mass of a terrestrial planet must be greater than 0.3 M$_\oplus$ to enable the planet to retain its atmosphere and sustain tectonic activity. However, the planetary disks surrounding low-mass stars may not be massive enough to form planets in this range, with the fraction even smaller for disks around stars in the 0.5-0.8 M$_\odot$ range. Water delivery may be inefficient in these low-mass disks, resulting in habitable-zone terrestrial planets that are both too small and dry to remain habitable around low-mass stars \citep{Raymond2007}, particularly given their increased luminosity during the period of planet formation \citep{Lissauer2007}, and lower surface density of solids \citep{Ciesla2015}. Indeed, one of the only ways to form volatile-rich Earth-sized planets around M-dwarf stars may be through the evaporation of the H/He envelopes of mini-Neptunes by XUV radiation followed by inward migration toward the parent star \citep{Luger2015a}. However, given the possibility of larger-sized dust grains in the protoplanetary disk, the lower resulting dust opacity could result in a snow line that is closer in to the star, increasing water delivery \citep{Mulders2015b}. Considering the higher planet-occurrence rate for M-dwarf stars \citep{Howard2012, Mulders2015a}, water-rich terrestrial planets could dominate around M-dwarf stars. 

Statistical studies have also found a larger cumulative mass of planets, in addition to a greater occurrence of small (1-2.8 R$_\oplus$) planets, around M-dwarf stars compared to more massive host stars \citep{Mulders2015c}, as shown in Figure 2. This counters earlier theories of the relationship between protostellar disk mass and resultant planet formation, indicating that a different mechanism may govern the formation of small, volatile-rich planets. One such possible mechanism involves a significant amount of planet-forming material migrating inwards from more distant regions of a protostellar disk, thereby providing enough mass to build greater numbers of small planets close to their stars, and larger cumulative planetary masses for M-dwarf host stars \citep{Mulders2015c}. And the fraction of low-mass stars with stellar companions ($\sim$23.5\%, \citealp{WardDuong2015}) is lower than solar-type stars ($>$40\%, \citealp{Raghavan2010}), which may reduce the likelihood of protoplanetary disk truncation in M-dwarf systems \citep{Artymowicz1994}.  

Recent research suggests that not only are small planets more common around smaller, lower mass stars \citep{Howard2012, Mulders2015a, Mulders2015c}, but surveys have found that these smaller, Earth-sized planets are typically not alone in their systems. Statistical data have identified a higher occurrence rate of planets orbiting M-dwarf stars than earlier-type stars \citep{Swift2013}. Indeed, about one-third of \textit{Kepler}'s planet candidates reside in systems with planetary companions \citep{Lissauer2012b}. This suggests that multiple-planet systems are a major planetary population that may be the primary type of environment targeted in the search for another habitable planet like the Earth. In the last few years, many systems of multiple planets orbiting M-dwarf stars have been discovered \citep{Anglada2013, Quintana2014, Rowe2014, Torres2015, Crossfield2015, Barclay2015}. The smallest habitable-zone planet discovered to date\textemdash Kepler-186f\textemdash was found orbiting near the outer edge of its star's habitable zone in a five-planet system \citep{Quintana2014}. Kepler-186f is only 10\% larger than the Earth, and although its mass, density, and bulk composition are unknown, recent work suggests that it is small enough to be rocky \citep{Rogers2015}.     
 
The gravitational effects that can occur in systems where multiple planets orbit a central star are of particular interest to the subject of M-dwarf planet habitability, given that the large majority of multiple-planet systems around M-dwarf stars exist in close, dynamically packed orbits \citep{Fang2013}. Some systems, such as the Kepler-32 five-planet system, have been found orbiting their stars at a fraction of the distance between Mercury and the Sun \citep{Swift2013}. As mentioned in Section 4, these gravitational interactions can have significant effects on dynamical stability, the evolution of planets' orbital parameters such as eccentricity and semi-major axis, and annually-averaged instellation and climate. However, the range of dynamically stable eccentricities possible for potentially habitable planets residing in multiple-planet systems can be fairly wide \citep{Shields2016a}. 

As more multiple-planet systems are discovered in the coming years by missions such as the Transiting Exoplanet Survey Satellite (TESS, \citealp{Ricker2009, Ricker2014}), discussions to constrain the possible climates and potential habitability of such systems will need to include input from \textit{n}-body simulations that have evolved the orbits of these planets over long timescales. This approach will confirm that orbital stability is maintained over time periods significant to biological evolution. Individual case studies that combine both orbital dynamics and climate model simulations (see e.g., \citealp{Shields2016a}) will prove critical to target selection for future missions that hope to characterize their atmospheres, such as the James Webb Space Telescope \citep{Gardner2006} and later missions.

\section{Life on an M-dwarf Planet: Extreme Life Effects}\label{sec6}
Life as we know it on Earth requires three things: Liquid water; a suitable environment for the formation of organic molecules from bioessential elements (sulfur, phosphorous, oxygen, nitrogen, carbon, hydrogen); and an energy source, whether stellar or chemical \citep{DesMarais2008}. A comprehensive discussion of the theoretical considerations that are relevant to the potential habitability of M-dwarf planets naturally leads to the exploration of the kinds of life that may be expected to emerge and evolve on these worlds. Given the possibilities of extreme temperatures on the day and night sides of a synchronously-rotating M-dwarf planet, high stellar activity over extended evolutionary timescales, and low water concentration, it is important to examine the environmental limits of life as observed thus far on the only habitable planet currently known\textemdash the Earth. As DNA, the fundamental information molecule for life on Earth, is particularly susceptible to certain environmental extremes \citep{Rothschild2001}, it is crucial to understand how organisms survive and grow in different environments. This understanding will enable a more accurate assessment of the likelihood of an origin of life on an M-dwarf planet given the different climate conditions possible on these worlds.

Here we summarize research done to identify and characterize life in extreme environments on the Earth. We also address the likelihood and availability of photosynthetic processes on M-dwarf planets, based on work carried out in recent years that included the specific spectral properties of M-dwarf stars. Finally, we discuss implications of these results for the survival and growth of life elsewhere in the universe in the context of environmental and climate conditions possible on M-dwarf planets. 

\subsection{Extremophilic Life}
The term ``extremophiles" is assigned to organisms discovered to thrive in environments that we judge to be extreme. These environments can include physical extremes of temperature, pressure, or radiation, as well as geochemical extremes of pH, dessication, salinity, oxygen species, or redox potential (the tendency of a chemical species to acquire electrons, thereby becoming ``reduced").  

Thriving organisms have been found within a wide range of different habitats, including the steaming hot springs of Yellowstone National Park \citep{Segerer1993}, the hot, arid regions of the Atacama Desert in Chile \citep{McKay2003}, subglacial lakes in East Antarctica \citep{Bulat2004, Bulat2011} and colder, arctic wintertime sea ice at temperatures well below 0$^\circ$C \citep{Junge2004}, and the hypersaline waters of the Red Sea \citep{Krumbein2004}. Some organisms thrive in environments with multiple extremes, amidst both extremely acidic and hot conditions, for example. The organism \textit{Sulfolobus acidocaldarius}, perhaps one of the best studied extremophilic members of the taxonomic domain of Archaea\footnotemark{}, was found in a Yellowstone National Park Geyser amidst a pH of 3 (highly acidic) and an average temperature of 80$^\circ$C \citep{Rothschild2001}. Hydrostatic pressures at the bottom of the ocean allow organisms to thrive in even higher temperature surroundings at hydrothermal vents \citep{Baross1985, Segerer1993, Pledger1994, Prieur1995} and sulfide chimneys or ``black smokers" \citep{Baross1983, Deming1993, Schrenk2003}, and in domains that are nutrient- \citep{Hirsch1986, Hoehler2013} and oxygen- \citep{Stetter1984, Jorgensen2007} poor. \footnotetext{Archaea, along with the Bacteria domain, are single-celled prokaryotic microorganisms that lack a cell nucleus.}

The extreme of temperature is one of the most relevant factors to consider in the discussion of suitable planetary conditions for life. The solubility of gases decreases with increasing temperature, therefore some water-based organisms that require O$_2$ and CO$_2$ can encounter difficulties at high temperatures \citep{Rothschild2001}. However, the proteins, nucleic acids and membranes of hyperthermophiles are known to be stable at least to 122 $^\circ$C (under pressure; \citealp{Takai2008}). 

The high-temperature extreme for the maximum growth of life on Earth is found among the organisms defined as hyperthermophilic (``heat-loving"), with growth above a temperature of 80$^\circ$C. The most hyperthermophilic organisms belong to the archaea domain, some of which are capable of growth at temperatures as high as 113$^\circ$C \citep{Blochl1997, Rothschild2001, DesMarais2008}, with the current upper limit being 122$^\circ$C under high-pressure conditions \citep{Takai2008}. The upper temperature limit for single-celled eukaryotes (organisms with a cell nucleus) is much lower, at 60$^\circ$C, and for multicellular organisms, even lower (e.g., vascular plants), at 48$^\circ$C. Select phototrophic (light-harvesting) bacteria, as well as other metabolic types of bacteria, are also thermophilic \citep{Rothschild2001}, and even hyperthermophilic \citep{Takekawa2015} . 

The cryosphere on Earth offers many available habitats for psychrophiles (``cold-loving" organisms) to flourish. Psychrophilic organisms demonstrate optimal growth at T $<$ 15$^\circ$C \citep{Morita1975}, and are capable of growth at temperatures well below 0$^\circ$C; the current lower temperature record for growth is $-15^\circ$C held by a bacterium from permafrost \citep{Mykytczuk2013}. Microbial organisms (algae, protozoa, bacteria, archaea) reside in sea ice \citep{Staley1999, Collins2010}, within its liquid brine inclusions \citep{Junge2001}. Because cold temperatures reduce protein fluidity, mechanisms that allow for the increased flexibility of membranes have been shown to improve structure and activity at low temperatures \citep{Aghajari1998}. In fact, the excretion of extracellular polymeric substances (EPS) by organisms in cold environments has been shown to protect cells from freezing, and allow prolonged function across a range of temperature changes in sea ice \citep{Ewert2013}. EPS has been known for a long time \citep{McLean1918}, but its full capability as an environmental buffer in the cold has only recently been studied \citep{Deming2002, Krembs2011, Ewert2013}. 

The deep sea is a multiple-extreme environment of both high pressures and low temperatures, except at hydrothermal vents, where temperatures can reach 460$^\circ$C \citep{Bischoff1984, Koschinsky2008} and high pressures sustain the water in liquid form. Indeed, pressure has been shown to stabilize proteins at temperatures well above typical denaturing temperatures (to 111$^\circ$C; \citealp{Summit1998}), extending the maximum temperatures for survival in the deep-sea hydrothermal vent environment \citep{Pledger1994, Zeng2009}. Some hyperthermophiles from deep-sea vents have an absolute requirement for high pressure in order to grow \citep{Zeng2009}.

Though life on Earth uses a diversity of metabolisms, water seems to be a universal requirement \citep{DesMarais2008}\textemdash for chemical bonding, and as a solvent for chemical reactions. Therefore, it is liquid water that currently drives the search for life elsewhere. Since there is a wide range of temperatures over which water can remain liquid and form hydrogen bonds, its limited supply also constitutes an extreme environment that could severely impact the survival and growth of life. Given the recent results suggesting that habitable-zone M-dwarf planets may be dry\textemdash either due to low protostellar disk mass \citep{Raymond2007}, or extended XUV radiation input over as long as 1 Gyr \citep{Luger2015b}\textemdash and land planets are both less likely to have saturated atmospheres prone to hydrogen escape and less susceptible to freezing at large distances from their star \citep{Abe2011}, the influence of dry, dessicated environments on habitability and life is of particular importance. 

Without an abundant supply of the water molecule, ``anhydrobiosis" can occur, where an organism becomes so dehydrated at the intracellular level that metabolic activity virtually ceases. This dehydration can cause death as a result of ensuing effects, including the denaturing of lipids, proteins, and nucleic acids, and the production of reactive oxygen species during dessication as a byproduct of exposure to harsh stellar radiation \citep{Cox1993, Dose1994, Dose1995, Rothschild2001}.  

The various mechanisms life employs amidst conditions of low water availability on the Earth are particularly applicable to M-dwarf planets, given their potential to be volatile poor \citep{Lissauer2007, Raymond2007, Luger2015b}. Hot dry regions on Earth such as the Atacama desert in Chile, and the coldest, driest regions in Antarctica have provided a suitable lab for the study of how organisms survive by avoiding dessication. The primary forms of life that have been found in these regions include cyanobacteria, algae, and fungi. Since water is of limited supply, life is often found in microbial mat crustal communities, within sandstone very close to the surface \citep{Evans1999}. In Antarctica these nutrient-limited photosynthetic communities \citep{Thielen1999} have become adapted to withstand prolonged intervals of darkness, using melting snow as a temporary water source in otherwise arid conditions \citep{Friedmann1982}. Bacteria can increase their osmotic pressure in response to dry conditions \citep{Yancey1982}, allowing water to flow from areas of low solute concentration to high solute concentration, protecting the cytoplasm (cell interior) from dehydration, and stabilizing proteins \citep{Potts1994}.

Microbial organisms have been found in hypersaline deep-sea environments, salt lakes, and evaporite ponds. Halophilic (``salt-loving") archaea are particularly suited to such environments, but certain types of bacteria, cyanobacteria, algae, diatoms, and fungi can also tolerate high salt content \citep{Rothschild2001}. Microorganisms persist in fluid inclusions in salt crystals \citep{Norton1988}, potentially surviving for timescales on the order of millions of years \citep{Vreeland2000}, and use atmospheric carbon dioxide and nitrogen to fix organic compounds within gypsum halite crusts \citep{Rothschild1994}. Such modalities could be especially useful on a dry M-dwarf planet.

Environments of extreme temperature and/or salinity may also be accompanied by additional extremes of pH. Microorganisms that are capable of survival and growth in highly acidic or highly alkaline conditions do so through the use of either active mechanisms (such as proton uptake) or passive mechanisms (altered permeability characteristics or surface charge properties, for example) to regulate the pH of their cell cytoplasm, keeping it comparable to that of organisms that do not live at extreme pH  \citep{Rothschild2001}. Such features of highly-adaptable organisms would be valuable within an environment subject to changes in water content and salinity.  

The most apparent obstacle to surface life on an M-dwarf planet is the intense XUV radiation emitted by their host stars \citep{Hawley1991, Segura2010, Luger2015b}. Radiation\textemdash carried by either particles, such as protons, neutrons, and alpha particles, or as electromagnetic waves, such as X-rays, Gamma and UV rays\textemdash can cause significant damage to biology. UV radiation in particular, has been known to negatively impact motility and prevent photosynthesis, and can seriously damage DNA, largely through the creation of heavily reactive oxygen species \citep{Rothschild2001}. Studies in laboratory and space environments have revealed that certain Earth organisms are able to tolerate large doses of high-energy radiation, by employing strategies such as the production of antioxidants and enzymes that aid in detoxification, as well as repair mechanisms, favorable assembly configurations \citep{Rothschild1999, Rothschild2001}, and overlying dust layers \citep{Mancinelli2000}. A well-known example of a highly studied radiation-tolerant organism is \textit{Deinococcus radiodurans}, which tolerates extreme levels of radiation using a mechanism that allows it to repair fragmented DNA and prevent the loss of encoded information \citep{Battista1997,Battista1998, Diaz2006}. Its resistance to high levels of ionizing gamma radiation has been linked to its dessication tolerance \citep{Battista1997}.

Given that M-dwarf planets could build up significant O$_2$-rich atmospheres \citep{Luger2015b}, the photochemical production of the hydroxyl radical (OH$^-$)\textemdash a reduced form of O$_2$\textemdash and hydrogen peroxide (H$_2$O$_2$) is likely, and these highly reactive oxygen species can exacerbate radiation-induced DNA damage \citep{Tyrell1991}. O$_2$-rich M-dwarf planets could already be dessicated however, and void of life when we observe them today \citep{Luger2015b}. Yet given the lengthy extended PMS phase for M-dwarf stars, there is likely to be some lag time between the saturation of a stratosphere and the dessication of the entire planet. Regardless, the production of highly reactive oxygen species is likely to hasten existing biological damage induced as a result of intense XUV irradiation on an M-dwarf planet. 

\subsection{Photosynthesis on M-dwarf Planets}
The process of photosynthesis is used by plants to convert CO$_2$ and light energy (plus an electron donor, such as H$_2$O) to carbohydrates and water (plus an oxidized electron donor). A prevailing theory connects the origin of photosynthesis to the evolution of organisms from a dependence on deep-water hydrothermal vent environments to an adaptation to shallower waters, where the harvesting of solar energy was possible \citep{Nisbet1995, Nisbet1999, DesMarais2000}. The first photosynthesizers were likely anoxygenic purple bacteria and green sulfur bacteria that used reduced species other than water, such as H$_2$S \citep{Olson2006}. Oxygenic photosynthesis was first carried out on Earth with the emergence of cyanobacteria\textemdash the first organisms to generate oxygen as a byproduct from water \citep{Ward2015}. This development resulted in an increase in the O$_2$ concentration in the atmosphere. The period of transition in the Earth's atmosphere from a reduced to an oxidized state comprises what is termed the Great Oxygenation Event ($\sim$2.4-2.0 Ga). With the emergence and evolution of plants onto land, O$_2$ levels continued to rise, peaking during the Carboniferous period ($\sim$360-300 Ma\protect\footnotemark{}), and eventually stabilizing at the $\sim$20\% level we see today \citep{Holland2006}.
\footnotetext{``Ga" and ``Ma" are abbreviations for ``billion years before present" and ``million years before present", respectively.}

In plants, it is the pigment chlorophyll \textit{a} (Chl \textit{a}) that is responsible for absorbing incoming solar radiation at selected wavelengths, and performing charge separation to gather electrons from an electron donor, such as H$_2$S or H$_2$O. The range of the spectrum where oxygenic photosynthesis occurs on Earth is largely restricted to the 400-700-nm range, called ``photosynthetically active radiation" (PAR). Chl \textit{a} uses this PAR, which offers abundant energy to carry out the reactions of photosynthesis. Specifically, Chl \textit{a} (and other accessory pigments that vary by organism) absorbs strongly in the blue and also in the red region of the visible spectrum (Figure 10). The lower absorption coefficient in the green range of the spectrum is responsible for the higher reflectance of plants in this range, and their resulting green appearance to the human eye. 

\linespread{1.15}
\begin{figure}
\begin{center}
\includegraphics [scale=0.60]{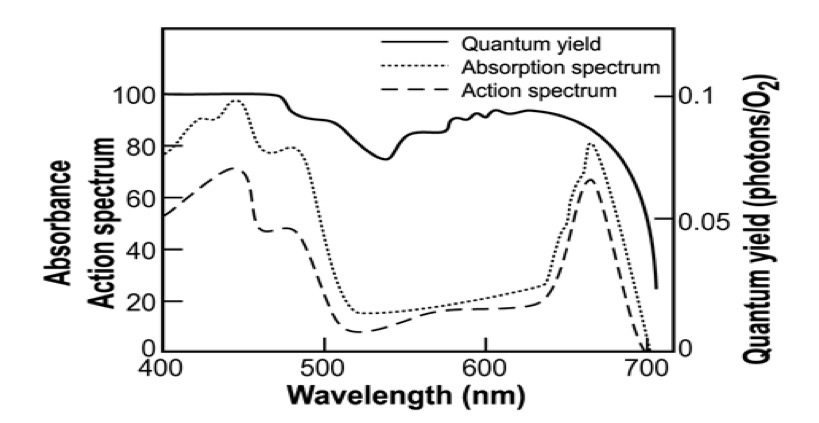}\\
\caption{Figure 2 from Gale and Wandel (\citeyear{Gale2015}), showing the absorption spectrum, as well as the action spectrum and quantum yield for chlorophyll while engaged in the process of oxygenic photosynthesis. Note the drop in the absorption coefficient in the green range of the spectrum (495-570 nm), and the sharp drop at $\sim$700 nm. This figure is reproduced from \citet{Gale2015} with permission from the authors and Cambridge University Press.} 
\label{Figure 10. }
\end{center}
\end{figure}

However, the range of PAR on the Earth constitutes $\sim$48\% of the total incoming solar radiation \citep{Gale2015}, with the rest of the incoming solar radiation at longer wavelengths. In fact, $\sim$50\% of the sun's radiation is emitted at wavelengths longer than 0.7 $\mu$m. Plant leaf absorption drops substantially at longer wavelengths, from about $\lambda$ $\ge$ 0.7 $\mu$m, which is responsible for the vegetation ``red edge" at $\sim$680-760 nm \citep{Gates1965, Seager2005, Kiang2007a}, as shown in Figure 11.

\linespread{1.15}
\begin{figure}
\begin{center}
\includegraphics [scale=0.60]{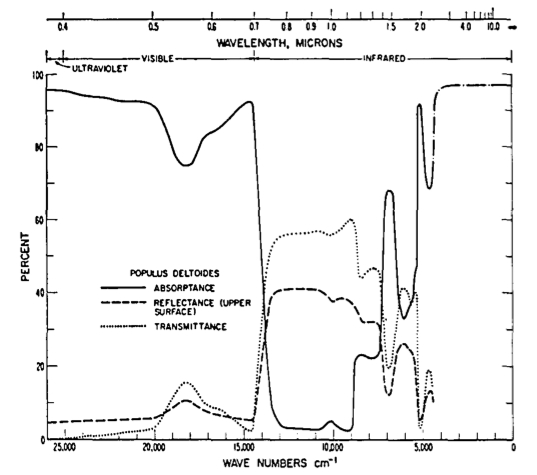}\\
\caption{Figure 3b from Gates \textit{et al.} (\citeyear{Gates1965}), showing the reflectance, transmittance, and absorptance as a function of wavelength for leaves from \textit{Populus deltoides}, a cottonwood tree with thin, light colored green leaves. Its absorption, which is very efficient at short wavelengths, drops substantially at 0.7 $\mu$m, and then increases again at $\sim$ 2.5 $\mu$m, where solar emission is low. This figure is reproduced from \citet{Gates1965} with permission from the authors and OSA.} 
\label{Figure 11. }
\end{center}
\end{figure}

If plants absorbed in the near-IR with the same efficiency as they do UV and visible radiation, the resultant heating would put the proteins within their structures at risk of denaturing. Beyond 2.5 $\mu$m, the sun emits far less strongly, therefore plant absorption efficiency increases significantly across this range \citep{Gates1965}.

On a world receiving a large portion of incoming stellar radiation in the near-IR from a small, dim host star, the lower availability of PAR photons has called into question the likelihood of photosynthesis to occur on M-dwarf planets. Although a number of photosynthesizing bacteria are known to contain pigments that absorb at IR wavelengths\textemdash certain genera of green sulfur and purple non-sulfur bacteria, for example\textemdash they do not use water as their electron donor, therefore they do not produce oxygen as a byproduct, and do not contribute to the build-up of a UV-shielding ozone layer \citep{Heath1999}. Under limited conditions for visible photons, it was questioned whether anoxygenic or oxygenic photosynthesis (in which water is the electron donor) would be viable on M-dwarf planets. 

Previous work addressing the spectral energy requirements and mechanisms of photosynthesis implied that in a photon-limited regime, photosynthesis can still occur, but with adaptations to the different light regime. A chief advantage for photosynthetic life on the sunlit side of a synchronously rotating planet was highlighted, as vegetation would receive more stellar insolation at the planet's non-varying substellar point over a 24-h period than over the same period on the more rapidly rotating Earth \citep{Heath1999}. Additionally, it was proposed over a decade ago that oxygenic photosynthesis using different ranges of the electromagnetic spectrum other than visible wavelengths could occur on other planets orbiting stars of different spectral types, including longer-wavelength photons for planets orbiting M-dwarf stars \citep{Wolstencroft2002}. Surface vegetation that can absorb more photons would simply have an edge over vegetation with more stringent photon requirements.

It remains an area of research to understand the processes that govern the wavelengths harvested for photosynthesis. Kiang \textit{et al.} (\citeyear{Kiang2007b}) proposed that photosynthesis on planets orbiting other stars should evolve to occur using that type of light transmitted to the surface with the greatest photon flux densities, and this is largely demonstrated by Earth phototrophs, which utilize a combination of accessory pigments as well as the primary chlorophyll or bacteriochlorophyll to tune spectrally for the available incident light. For example, while the sun's energy density peaks in the range of 450 nm, its photon flux density peaks in a longer wavelength range of the spectrum, 572-584 nm \citep{Kiang2007b}. And as Kiang \textit{et al.} (\citeyear{Kiang2007a}) showed, the photon flux density that actually transmits through the atmosphere to reach the surface peaks at an even longer range of the spectrum, 670-680 nm. It is these photons that are most available to surface vegetation to harvest for the reactions of photosynthesis. 

In general, simple thermodynamics favor that the harvested light must occur at shorter wavelengths than the ``trap" or band gap wavelength at which charge separation occurs, and the peak photon flux density might occur anywhere within that range.  However, there are cases in which organisms have pigments harvesting light at longer wavelengths than the trap, such that the photon energy must climb an energy hill to be used.  Explaining this phenomenon requires further research into the role of quantum tunneling in photosynthesis. 

Kiang \textit{et al.} (\citeyear{Kiang2007b}) calculated photon flux densities reaching the surface on land and underwater (Table 1), and found that the photon surface flux density on planets orbiting M-dwarf stars would peak in particular near-IR wavebands at 0.93-1.1 $\mu$m, 1.1-1.4 $\mu$m, 1.5-1.8 $\mu$m, and 1.8-2.5 $\mu$m. As water absorption bands coincide with the longer of these bands, underwater organisms would have access to only those bands shortward of 1.1-1.4 $\mu$m. 

\linespread{1.0}
\begin{table}[!htp] 
\caption{Incident photon flux densities at solar noon for a cloudless planet orbiting a G2V and an M5V host star, from Table 2 of Kiang \textit{et al.} (\citeyear{Kiang2007b}). The G2V planet (Earth around the Sun) assumes 1 PAL O$_2$. The last two columns show photon flux densities underwater at depths of 5 cm and 100 cm, respectively assuming low O$_2$ (O$_2$ $\times$ 10$^-5$). This table is reproduced from \citep{Kiang2007b} with permission from the authors and Mary Ann Liebert, Inc.} 
\vspace{2 mm}
\centering \begin{tabular}{c c c c c c} 
\hline\hline 
& \textit{G2V (Sun/Earth)} & \textit{M5V O$_2$ $\times$ 10$^-5$} & \textit{M5V (1 PAL)} & \textit{5 cm} & \textit{100 cm} \\  [0.5ex]
\hline
\hspace{10mm}UVB 280-315 nm & 0.018 & 0.000 & 0.000 & \textemdash &  \textemdash \\
\hspace{10mm}UVA 315-400 nm & 0.871 & 0.016 & 0.016 & \textemdash &  \textemdash \\
PAR & & & & &\\
\hspace{10mm}
\hspace{8mm}
400-700 nm & 11.0 & 1.5 & 1.5 & 1.4 & 1.1 \\
\hspace{21mm}
400-1,100 nm & 23.8 & 17.3 & 16.9 & 9.9 & 1.5 \\
\hspace{21mm}
400-1,400 nm & 28.6 & 25.7 & 24.9 & 10.2 & 1.5 \\
\hspace{21mm}
400-1,800 nm & 33.7 & 35.3 & 34.3 & 10.2 & 1.5 \\
\hspace{21mm}
400-2,500 nm & 36.9 & 40.1 & 38.4 & 10.2 & 1.5 \\
\hspace{31mm}
Peak photon flux & 668.5 & 1,042.8 & 1,042.8 & 1,073.2 & 1,073.2\\
[1ex]
\hline 
\end{tabular} 
\label{table:nonlin} 
\end{table}

Due to the lower number of available photons typically constituting PAR for M-dwarf stars, photosynthesis could be as much as ten times less productive on M-dwarf planets than on Earth at visible wavelengths. However, if available photons for photosynthesis extend to 1.1 $\mu$m, anoxygenic photosynthesis could be carried out on M-dwarf planets with equal productivity compared to Earth, if not limited by electron donors \citep{Kiang2007b}. Indeed, the competitive edge goes to organisms that can take advantage of more resources, such as light across a broad range of wavelengths.

\subsection{Where Life Can Survive on M-dwarf Planets}
Given the diversity of mechanisms employed by many different species on the Earth to facilitate survival and growth within environmental extremes, it seems plausible that in the event of similar or even more limiting conditions, surface life may consider analogous modes to sustain metabolic processes on M-dwarf planets. 

Experiments investigating the tolerance of Earth organisms to extreme conditions have been carried out with an eye toward an application to the environments of the early Earth \citep{Cnossen2007, Westall2011, Grosch2015}, solar system planets like Mars \citep{Navarro-Gonzalez2003, Cockell2004, Diaz2006, delaVega2007, Dartnell2010}, the moons of the outer solar system such as Jupiter's moon Europa \citep{Chyba2000, Chyba2002, Deming2002, Marion2003, Bulat2004, Bulat2011, Kimura2015, Noell2015}, and interplanetary space \citep{Paulino-Lima2010}. However, given the varied factors that may influence the climate and surface environment of M-dwarf planets, much of what has been learned about the mechanisms employed by extremophilic life on Earth to not only survive but thrive has implications for the likelihood of life to grow and evolve on planets in close-in orbits around these small, cool stars. 

DNA is particularly susceptible to extremes of radiation, high temperature, oxidative state, and dessication \citep{Rothschild2001}. Additionally, low temperature \citep{Mazur1980, Deming2002, Diaz2006}, and low levels of PAR \citep{Kiang2007a, Kiang2007b} are of direct interest to the discussion of available habitats for life on M-dwarf planets, with the possible secondary influences of pressure, high salinity and extremes of pH.  

On a synchronously-rotating M-dwarf planet with a thin atmosphere, insufficient heat distribution would result in a large temperature contrast between the day and night sides of the planet, with the day side blisteringly hot, while the night side could be completely frozen. The ability of surface life to thrive in cold or even icy conditions, as well as high-temperature environments, could be particularly relevant. Depending on the atmospheric concentration and low temperatures, the surface on the night side could be covered in a combination of both H$_2$O and CO$_2$ ice. In this case, hyperthermophilic life would necessarily be predisposed toward life on the day side, providing temperatures did not exceed the current life limit of 122$^\circ$C (under pressure; \citealp{Takai2008}). This environment would also be preferred for photosynthetic organisms with high-temperature tolerance, given that they would have access to high amounts of incoming stellar radiation. On the cold night side, at temperatures approaching or above 0$^\circ$C, psychrophilic or psychrotolerant life could easily exist. Assuming the planet has an ocean that freezes over on the night side, psychrophilic algae, bacteria and archaea could live in sea ice within brine inclusions as on the Earth, and any EPS-like compounds would provide protection against ice-crystal formation near enough to damage (puncture) cellular membranes.    

M-dwarf planets might be dry worlds (e.g., \citealp{Raymond2007}), predominantly covered by deserts and evaporite deposits, especially if temperatures become so high on the day side that oceans evaporate \citep{Luger2015b}. Given that abundant microbial mat communities have been found in hot and cold desert environments, as well as living within fluid inclusions in salt crusts and other mineral halite crustal communities on Earth, similar microorganisms might be capable of survival and growth in such conditions elsewhere, provided that some amount of water is available. 
The intense XUV activity emitted by M dwarfs early in their lifetimes may have played a significant role in the evolution of life on orbiting planets. While on the Earth life evolved to spread to above-water surroundings on land, on an M-dwarf planet such life may have remained concentrated in deep-ocean environments. Underwater organisms would be both shielded from potentially harmful UV flares emitted by the parent star, and could also have access to sufficient light for photosynthesis \citep{Kiang2007b}. Given phylogenetic and other evidence of a thermophilic origin of life on Earth \citep{Baross1985, Pace1997}, it would be a natural extension to envision a similar emergence of life elsewhere in the universe, particularly in an environment with an inherent benefit (radiation avoidance) to sub-aqueous development and evolution. Life may have existed within hydrothermal vents under conditions of high temperatures and pressures, while other life in the deep ocean could develop low-temperature high-pressure tolerances. 

As O$_2$-rich atmospheres are a possibility on M-dwarf planets \citep{Luger2015b}, the danger to biology of reactive oxygen species produced in the atmosphere is relevant, as these species could exacerbate the damage done to DNA by XUV radiation \citep{Tyrell1991}. Provided that these planets are not completely dessicated (there is still some water available to sustain life, if on a smaller scale), life could possibly avoid some of this oxidative damage if there is an O$_2$ sink on the planet, such as continental weathering or volcanic outgassing of reducing gases (H$_2$, H$_2$S, CH$_4$), as well as plate tectonics \citep{Luger2015b}. Such a sink could prevent the build-up of reactive oxygen species, thereby mitigating against an additional complication for life struggling to survive within a dry, dessicated environment. 

The O$_2$-rich environments possible on M-dwarf planets could set the stage for the formation of complex life given the abundant supply of energy required for aerobic metabolism, which is employed by the majority of multicellular life on the Earth \citep{Catling2005}. But given that more complex, multicellular life forms have a narrower range of temperature tolerance (see Section 6.1), the varied possible extremes of temperature, along with the extremes of radiation exposure and water availability could be even more challenging for advanced life forms (such as humans on the Earth).  However, as has been discussed, synchronously rotating M-dwarf planets with thicker atmospheres would likely have smaller temperature contrasts between their day and night sides, and M-dwarf planets with Earth-like rotation rates and atmospheres may be less susceptible to global-scale glaciation than their counterparts orbiting hotter, more luminous stars. These results imply that M-dwarf planetary environments also have the potential to be amenable for advanced life forms, depending on planetary atmospheric composition and rotation rate, among other factors.

\section{Conclusions}\label{sec7}
The prospects for life on planets orbiting M-dwarf stars have been the subject of extensive debate ever since the first planetary mass companion was discovered orbiting another main-sequence star \citep{Mayor1995}. The intense and prolonged stellar activity of these long-lived stars, as well as the gravitational effects between the stars and close orbiting planets have generated particular concern about planetary habitability. However, largely due to theoretical modeling in multiple dimensions, we have now quantified a number of these effects.  While some characteristics of the M-dwarf stellar and planetary environment are still concerning, and more work remains to be done in the future to fully constrain their impact on habitability, some of these effects now indicate advantages for habitability, and serve as reasons to prioritize M-dwarf planets as targets in the search for the next planet where life exists. 

The past decade bore witness to NASA's \emph{Kepler} mission \citep{Borucki2006}, which drastically increased the number of confirmed exoplanets discovered. Of greatest significance among these planets is the discovery of numerous small planets close in size to the Earth orbiting in their host stars' habitable zones. Many of these habitable-zone planets orbit M dwarfs. Whereas 10 years ago we knew of no planets smaller than $\sim$4 R$_\oplus$, we now have sufficient data to make statistical arguments about the number of Earth-sized planets orbiting in the habitable zones of M-dwarf stars \citep{Kopparapu2013c, Dressing2015}, and it has become clear that the M-dwarf stellar environment may be the primary type of environment that we look at in our search for another habitable planet like the Earth. 

The ultimate negative effect of stellar flares from M dwarfs on ozone column density may be limited (unless proton fluxes are included), and the strength of a planet's magnetic field could mitigate against ozone loss, as well as atmospheric erosion by stellar winds. However, the strength of the magnetospheres of M-dwarf planets may be precarious. Additionally, the extended PMS phases of M dwarfs could present significant challenges for surface oceans and life, and possibly produce abiotic O$_2$-rich atmospheres on long dessicated worlds. However, such effects are a function of planetary orbital distance and stellar mass, and could be offset by O$_2$ sinks \citep{Luger2015b}.

While the prospect of synchronously-rotating M-dwarf planets has historically been a source of concern, significant research in this area has caused widespread opinion on the effect of this spin state on planetary habitability to shift. If a planet is synchronously rotating and has a dense enough atmosphere, sufficient heat distribution could prevent atmospheric collapse on the night side of the planet \citep{Haberle1996, Joshi1997, Wordsworth2015}. Synchronous rotation could even improve habitable surface conditions on planets orbiting at the inner edge of their stars' habitable zones, and extend the habitable zone for M-dwarf stars \citep{Yang2013}. 

Theoretical modeling studies have also provided a deeper understanding of the effect on climate of the interaction between an M-dwarf host star's spectrum and an orbiting planet's atmosphere and surface. Much of the large fraction of near-IR radiation emitted by M dwarfs would be absorbed by CO$_2$ and H$_2$O (if present in a planet's atmosphere), and water ice and snow on a planet's surface. As a result, M-dwarf planets are more resistant to global-scale glaciation, and may have more stable climates over long timescales, which could be advantageous for biological evolution \citep{Shields2013, Shields2014}. And photosynthesis, a process that is essential for a large fraction of life on the Earth, is no longer held to be an implausible mechanism to occur on M-dwarf planets. While perhaps less productive than on planets orbiting stars with more higher-energy photons available, M-dwarf planets could be capable of anoxygenic photosynthesis at the same level as on Earth if available photons extend to long-enough wavelengths \citep{Kiang2007b}. There are examples of life on Earth with peak absorbance at near-IR wavelengths \citep{Scalo2007}, and carrying out photosynthesis in conditions of low light intensity \citep{Beatty2005} . 

The launch of TESS \citep{Ricker2009, Ricker2014} will facilitate the discovery of thousands of exoplanets orbiting the nearest stars in the solar neighborhood. Given that most of these stars will be M dwarfs, many of the planets found by TESS will orbit these small, cool stars. Those planets found to orbit in their stars' habitable zones will be of particular interest, and will emerge as priority targets to follow up on with the James Webb Space Telescope \citep{Gardner2006} and later missions, to characterize their atmospheres. With a deepening understanding of the effects of the unique environments of M-dwarf stars on planetary climate and habitability, quantifying the probability of suitable conditions for life on the first M-dwarf planets discovered by TESS is within reach. This will allow the generation of a list of high-priority planets to target for follow-up spectroscopy to identify biosignature gases in their atmospheres. Armed with this list, the next generation of large-aperture mirror space telescope missions, with star shades or coronagraphs in tow, will know exactly where to point their gaze.

\section*{Acknowledgments}
This material is based upon work supported by the National Science Foundation under Award No. 1401554, by a University of California President's Postdoctoral Fellowship, and by the Juan Carlos Torres Fellowship at MIT. AS wishes to thank Rory Barnes, Jody Deming, Nancy Kiang, and Victoria Meadows for reading sections of this manuscript and providing essential feedback. SB gratefully acknowledges James Davenport as ever, for his exhaustive knowledge of low-mass star literature.

\pagebreak




\end{document}